\documentclass[prc,aps,superscriptaddress,twocolumn,showpacs,showkeys,amsmath,amssymb,nofootinbib]{revtex4-1}
\usepackage{graphicx}
\usepackage{dcolumn}
\usepackage{bm}

\begin{document}

\title{Description of Pairing correlation in Many-Body finite systems with density functional theory}

\author{Guillaume Hupin}
\affiliation{Grand Acc\'el\'erateur National d\textquoteright Ions Lourds, Boulevard Henri Becquerel, Bo\^ite Postale 55027, F-14076 Caen CEDEX 5, France}

\author{Denis Lacroix}
\affiliation{Grand Acc\'el\'erateur National d\textquoteright Ions Lourds, Boulevard Henri Becquerel, Bo\^ite Postale 55027, F-14076 Caen CEDEX 5, France}


\begin{abstract}
Different steps leading to the new functional for pairing based on natural orbitals and occupancies proposed in ref. [D. Lacroix and G. Hupin, arXiv:1003.2860] are carefully analyzed. Properties of quasi-particle states projected onto good particle number are first reviewed. These properties are used (i) to prove the existence of such a functional (ii) to provide an explicit functional through a $1/N$ expansion starting from the BCS approach (iii) to give a compact form of the functional summing up all orders in the expansion. The functional is benchmarked in the case of the picked fence pairing Hamiltonian where even and odd systems, using blocking technique are studied, at various particle number and coupling strength, with uniform and 
random single-particle level spacing. In all cases, a very good agreement is found with a deviation inferior to 
$1\%$ compared to the exact energy.
\end{abstract}
\maketitle

\section{Introduction}
\label{sec:intro}

\begin{figure}[htbq]
\begin{center}
\includegraphics[width = 8.5cm]{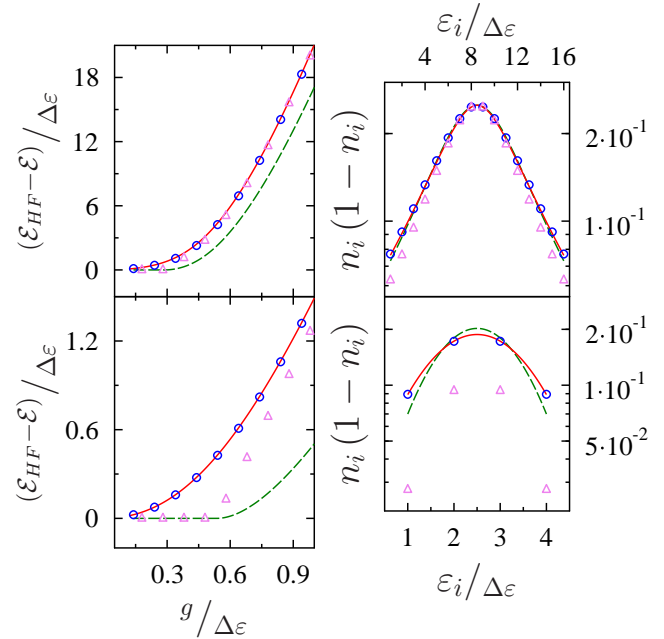}  
\end{center}
\caption{ \label{fig1:comp} (Color online) Exact condensation energy (red solid line) obtained for the picked fence pairing Hamiltonian as a function of the coupling strength for 16 (top) and 8 (bottom) particles. 
In both cases, the BCS (green dash line), the projected BCS  with a projection made before 
 (open blue circle)  or after the variation (open violet triangle) are also shown. In the right, occupation numbers of the different theories are plotted for $g/ \Delta \varepsilon=0.82$.}
\end{figure}

Nuclear systems \cite{Rin80,Bri05} or ultrasmall metallic grains \cite{Von01} offer the possibility to get insight in finite pairing correlations of systems with varying particle number. The introduction of a simple many-body wave packet ansatz more than 50 years ago by Bardeen, Cooper and Schrieffer (BCS) \cite{Bar57} was a major breakthrough for the understanding and the description of superconductivity. 
To illustrate the advantages and drawbacks of the BCS theory, in figure \ref{fig1:comp}, the condensation energy, i.e. the  
difference between the Hartree-Fock (HF) energy and the energy of the system obtained with BCS (dashed line) is 
compared to the exact result (solid line) for the picked fence pairing Hamiltonian (for details see section \ref{sec:application}) \cite{Ric65,Ric66,Ric66a}.   
One of the great advantage of the BCS or Hartree-Fock Bogolyubov (HFB) theory is the possibility, under the price to conserve 
particle number only in average, to grasp part of the correlation beyond the Hartree-Fock level while keeping the theory relatively simple. As can be seen from figure  \ref{fig1:comp}, 
the BCS prediction becomes closer to the expected result as the number of particle increases. Indeed, the BCS theory is shown to be exact in 
the thermodynamic limit. Besides these interesting aspects, BCS or HFB suffer from a threshold at low coupling. In fact, when the coupling strength is much smaller than the average level spacing between single-particle states, BCS identifies with HF while, 
in reality, correlations built up as soon as the two-body interaction is plugged in. {In addition, even above the threshold, part of the correlation are systematically missed.}

The BCS or HFB theories are nowadays standardly used in nuclear physics, for instance, within the Energy Density Functional (EDF)
approach \cite{Ben03,Sto07} leading to the so-called "Single-Reference"  (SR-EDF)  or "mean-field" level of EDF. These tools 
already provide a rather good reproduction of gross nuclear properties. For instance, masses can be estimated with a typical 
precision of 500-600 keV. Figure \ref{fig1:comp} however clearly points out that there is room for improving the BCS 
approach in finite size systems. In particular, part of the discrepancy stems from the use of a trial wave-function that is 
not an eigenstate of the particle number operator $\hat N$.  Starting from the BCS wave-packet, a new state with good particle number can 
be obtained using projection operator technique \cite{Bay60}. Within EDF, similarly to the restoration of angular momentum or 
{calculation including dynamical fluctuations associated to configuration mixing}, projection onto good particle number enters into 
the class of Multi-Reference EDF (MR-EDF). If the projection is made prior to the variation (Variation After Projection [VAP]), the
variational state directly becomes an eigenstate of $\hat N$.  Illustration of VAP condensation energy (open circles) is given in  figure 
\ref{fig1:comp} 
(see for instance \cite{Rod05}). Such an approach provides a very accurate description of pairing correlation at all coupling 
strengths and completely removes the BCS threshold problem. VAP still
remains rather involved numerically and a less efficient but simpler approach consists in projecting the state after the variation, 
the so-called Projection After Variation [PAV] (open triangles in figure \ref{fig1:comp}). Projection technique is becoming 
a popular tool in nuclear structure. However, recent studies have shown that projection aiming at restoring broken 
symmetries and/or more generally configuration mixing should be handled with care when combined with density functional 
theory\cite{Ang01,Dob07}, due to the possible appearance of jumps and/or divergences in the energy surface. These difficulties 
have been carefully analyzed in refs. \cite{Lac09a,Ben09,Dug09} and have been related to the self-interaction and self-pairing 
problem. By comparing theories starting from an Hamiltonian and an energy functional, a correction to the pathologies was proposed 
such that systematic calculation along the nuclear chart is now within reach. {These studies have clearly pointed out that specific 
aspects might appear due to the use of functional theories (see also \cite{Rob10,Dug10}) when MR-EDF is used. }

The EDF framework provides a unified framework not only for nuclear structure but also for nuclear 
dynamics and thermodynamics. While MR-EDF is a suitable tool for the former, due to its complexity, it can hardly 
be used in the latter cases.  The goal of the present work is to discuss a new approach to treat pairing 
where the projection effect is directly 
incorporated into the functional through specific dependencies on natural orbital occupancies.  Such an approach, 
directly written in the functional framework, avoids some ambiguities encountered in current EDF 
and is expected {to greatly simplify both PAV and} to be easily adapted to non-equilibrium evolution of finite temperature 
studies. Main aspects of the new functional theory have already been summarized in ref. \cite{Lac10}.
Here, we present a complete discussion of the different steps leading to the functional. Below, we first discuss the 
interest of using natural orbital based functionals. Then, mathematical properties of Projected BCS states that are 
used to propose the functional, are given. Finally, the new functional is applied to a specific pairing Hamiltonian 
either with equidistant or non-equidistant level spacing and benchmarked for any coupling strength and particle number.

\subsection{Functionals based on natural orbitals and occupancies}

The possibility to replace a many-body problem by a functional of the density matrix has been first proposed by Gilbert in ref. 
 \cite{Gil75} and is named Density Matrix Functional Theory (DMFT) or Reduced DMFT (RDMFT).  
 The Gilbert theorem is a generalization of the Hohenberg-Kohn theorem \cite{Hoh64} where the variational 
 quantity, i.e. the local density $\rho(\mathbf{r},\mathbf{r})$ is replaced by the full one-body density matrix (OBDM) $\rho(\mathbf{r},\mathbf{r}')$. Most often, the OBDM is first written
in the natural or canonical basis as $\gamma = \sum_i | \varphi_i \rangle n_i \langle \varphi_i |$. Here $n_i$ and $\{ 
| \varphi_i \rangle \}$ denote occupation numbers and natural orbitals respectively. Then, the initial many-body problem 
is replaced by the minimization of an energy functional
\begin{eqnarray}
{\cal F}[\{\varphi_i \}, \{n_i \}] &=& {\cal E} [\{\varphi_i \}, \{n_i \} ] - \mu \{ Tr({\hat N}\rho) -N \} \nonumber \\
&-&\sum_{ij} \lambda_{ij} (\langle \varphi_i | 
\varphi_j \rangle  - \delta_{ij}) ,
\label{eq:dmft}
\end{eqnarray} 
where the variation is made 
with respect to both single particle states $\varphi_i^*(\mathbf{r})$ and occupation numbers. 
The set of Lagrange multipliers $\mu$ and $\{ \lambda_{ij} \}$ are introduced to insure particle number conservation and 
orthogonality of the single-particle states. RDMFT has several advantages compared to standard 
Density Functional Theory (DFT). For instance, while Kohn-Sham single-particle states used  to construct the local density are not 
expected to have physical meaning, the non-local density $\gamma$ should match the exact one at the minimum. 
Accordingly, associated single-particle states and occupations identify with the one of the exact many-body state.
This is an important aspect of this theory. Indeed, DFT can only provide information on the energy. In RDMFT, not only 
the energy can be estimated but also any one-body operators. Similarly to Density Functional Theory, the main 
challenge is to find accurate functionals. 

Another interesting feature of this theory is its ability to describe aspects that are not adequately obtained 
at the DFT level, like reactions, atomization energy or the dissociation of small molecules. All these phenomena
have  their counterpart in nuclear physics. Nowadays, a sizeable effort is made to provide new accurate RDMFT 
functionals  and benchmark them on finite and infinite systems (see for instance \cite{Lat09} and refs. therein).

In this article, we focus on pairing. Let us first remark that current SR-EDF that account for pairing 
already share many aspects with RDMFT. Most nuclear SR-EDF used nowadays start from a functional that can be written as
\begin{eqnarray}
{\cal E}_{\rm SR} [\rho, C ]
& \equiv & {\cal E}^{\rho}
           + {\cal E}^{\rho\rho}
           + {\cal E}^{C}
           \nonumber \\
& = & \sum_{ij} t_{ij} \, \rho_{ji}
      + \frac{1}{2} \sum_{ijkl} \bar{v}^{\rho\rho}_{ijkl} \,
        \rho_{ki} \, \rho_{lj}
      \nonumber \\
&   & \phantom{ \sum_{ij} t_{ij} \, \rho_{ji}   }
      + \frac{1}{4} \sum_{ijkl} \bar{v}^{C}_{klij} \,
        C_{ij,kl}.
        \label{eq:functot}
\end{eqnarray} 
where $\bar v^{\rho \rho}$  and $\bar{v}^{C}$ denote effective two-body kernels respectively in the 
particle-hole and correlation channels. $C_{1,2}$ denotes the irreducible two-body correlation matrix defined as the difference 
between the two-body density and the antisymmetric product of one -body density matrix
(see for instance \cite{Lac04a}). 
To treat pairing correlations, a quasi-particle trial state, $| \phi_{QP} \rangle$, is considered, then 
the correlation matrix elements can be written in terms of the anomalous density $\kappa$ as $C_{ij,kl} = \kappa^*_{ij} \kappa_{kl}$ \cite{Rin80,Lac09a}.
In the natural orbital basis, the quasi-particle state can be expressed in a BCS form
\begin{eqnarray}
| \phi_{QP} \rangle  & = & \prod_{i} \left( 1 + x_i a^\dagger_i a^\dagger_{\bar i} \right) | 0 \rangle,
\label{eq:bcsstate}
\end{eqnarray}
where $| 0 \rangle$ corresponds to the particle vacuum while 
$\{ a^\dagger_i , a^\dagger_{\bar i} \}$ correspond to doubly degenerated canonical 
 states $\{\varphi_i , \varphi_{\bar i}\}$ with
 occupation probability $2 n_i$. The $x_i$ coefficients are connected to the 
 occupation numbers through  
 \begin{eqnarray}
|x_i|^2 = \frac{n_i}{(1-n_i)}.
\label{eq:nixibcs}
\end{eqnarray}
Accordingly, pairing energy reduces to:
\begin{eqnarray}
{\cal E}^C &=& \frac{1}{4} \sum_{i j} \bar v^{\kappa \kappa}_{i \bar i j \bar j} 
\sqrt{n_i (1-n_i)}  ~ \sqrt{n_j (1-n_j)}. \nonumber
\end{eqnarray}  
Noting in addition that both ${\cal E}^{\rho}$ and  ${\cal E}^{\rho\rho}$ can directly be written 
as a functional of $n_i$ and $\varphi_i $ through their dependence on the one-body density, we
see that current SR-EDF can indeed be interpreted as a mapping between 
the initial problem into a functional theory of $(\{\varphi_i \}, \{n_i \})$, i.e.   
\begin{eqnarray}
{\cal E}[\rho, C ] ~~~\rightarrow~~~ {\cal E} [\{\varphi_i \}, \{n_i \} ], \label{eq:mapping}
\end{eqnarray}
provided that the functional is written in the canonical basis. EDF based on quasi-particle states
have the shortcomings discussed in the introduction when combined with projection onto good particle 
number within MR-EDF. This is nowadays used in nuclear structure study. If the projection is made prior to 
the variation, such a projection is equivalent 
to consider a new trial wave-function, called hereafter generically Projected BCS (PBCS) state, of the form:
\begin{eqnarray}
| N \rangle  &\equiv& \hat P^N | \phi_{QP} \rangle  \propto \left( \sum_i x_i a^\dagger_i a^\dagger_{\bar i} \right)^N | 0 \rangle,
\label{eq:pbcsstate}
\end{eqnarray}  
where $\hat P^N$ is the projector on particle number $N$ (see for instance \cite{Bay60,Dob07,Dug09}). In the following, we will 
use the short notation $\Gamma^\dagger= \sum_{i} x_i b_i^\dagger $ with $b_i^\dagger = a^\dagger_i a^\dagger_{\bar i}$. 

When projection is made in EDF, the associated functional becomes much more complex to minimize. It should
however be noted that the functional is generally written in terms of the normal and anomalous density of the 
original quasi-particle state from which the projected state is constructed. Therefore, the occupation probabilities 
of the new trial state $| N \rangle$ do not appear directly \cite{She00}. Nevertheless, the occupation numbers of 
$| N \rangle$ can be estimated numerically. Illustration of occupation probabilities of the projected states are compared to the 
exact ones for the picket fence pairing Hamiltonian in figure \ref{fig1:comp}. Both VAP and PAV results as well 
as BCS case are displayed. It is first interesting to mention that, while the energy is improved in the PAV case, single-particle 
occupation numbers deviate more from the exact solution than the original BCS case. This is something 
to worry about since, when PAV is performed in EDF, expectation values of one-body operators are estimated.
In opposite, in the VAP method, occupation probabilities perfectly match the exact case for all particle number and pairing coupling 
strength. Therefore, we see that the use of projected state before the variation leads to a very good reproduction 
of both the ground state energy and the single-particle occupation numbers.

Having this in mind, in the following, we use the properties of PBCS state to provide a new functional 
for pairing directly based on the occupation numbers of the projected state. The state (\ref{eq:pbcsstate}) is used 
as a starting point where it is implicitly assumed that the orbitals are written in their canonical basis, namely the one which exhibits an explicit time reversal symmetry. 
In that case, the energy (\ref{eq:functot}) reduces to\footnote{Note that, correlation matrix elements should also 
appear in the particle-hole channel. Since the aim of the present article is to focus on pairing channel and since these components cancel out exactly in the example presented below, they are omitted here.}
\begin{eqnarray}
{\cal E}_{\rm SR} [\rho^N, C^N ]
& = & \sum_{i} t_{i} \, n^N_i
      + \frac{1}{2} \sum_{ij} \bar{v}^{\rho\rho}_{iijj} \,
        \rho^N_{ii} \rho^N_{jj} 
      \nonumber \\
&   & \phantom{ \sum_{ij} t_{ij} \, \rho_{ji}   }
      + \frac{1}{4} \sum_{ij} \bar{v}^{C}_{i\bar i j \bar j} \,
        C^N_{i\bar i,j \bar j},
        \label{eq:functotC}
\end{eqnarray}  
were $\rho^N_{ii} = n^N_i$ and $C^N_{i\bar i,j \bar j}$ now stand for the occupation and correlations associated 
with the projected state, i.e.
\begin{eqnarray}
n^N_i &=& \frac{\langle N  |a^\dagger_i a_i | N \rangle }{\langle  N | N\rangle }, ~~ 
C^N_{ij} = \frac{\langle N |b^\dagger_i b_j | N \rangle }{\langle N | N \rangle } - \delta_{ij} n^N_i n^N_j.\nonumber\\
\label{eq:nici0}
\end{eqnarray} 
Here, we have used the compact notation $C^N_{ij} \equiv  C^N_{i\bar i,j \bar j} $. 
In the following, we will omit the $N$ label to shorten notations keeping in mind that these quantities 
refer to the projected state.  
In order to do the mapping  (\ref{eq:mapping}), we are left with the challenge consisting in expressing 
the correlation $C_{ij}$ as a functional of $n_i$ as it can be easily done in the BCS or HFB case. 
But in the present work, we aim at accounting for the particle number conservation directly in the functional.


\section{Construction of functionals for pairing from a PBCS state} 

Here, some properties of projected states are first highlighted. 
These properties are then used as a guidance to construct 
the functional.   
Over years, interesting features of matrix elements entering in Eq. (\ref{eq:nici0}) have been derived. 
Some can eventually be deduced 
using the fact that the BCS state plays the role of PBCS state generating 
function \cite{Row91,Row01} and can be used, for instance, to minimize 
the energy directly written as a functional of the $\{ x_i \}$ parameters \cite{San08,San09}.
Proofs of some of the properties that are used below are first given.

\subsection{Definition of a class of operators, states and overlaps}

First, we start with a strategy similar to ref. \cite{Leg06}. A set of  
pair creation operators that omit one, two,... pairs of single-particle states is first introduced:
\begin{eqnarray}
\Gamma^\dagger (i) &=& \Gamma^{\dagger} - x_i b^\dagger_i \nonumber \\
\Gamma^\dagger (i,j) &=& \Gamma^{\dagger} - x_i b^\dagger_i - x_j b^\dagger_j, \nonumber \\
&\cdots& \label{eq:paircre}
\end{eqnarray} 
where indices $i$, $j$ refer to the removed pairs. 
In the following, $\Omega$ will denote the size of the single-particle Hilbert space 
From these operators, a corresponding set of states with a given particle number is defined:  
\begin{eqnarray}
\left\{
\begin{array} {ccc}
| K  \rangle &=&   c_K \left(\Gamma^\dagger \right)^K | - \rangle  \\
| K: i \rangle &=&  c_K \left(\Gamma^\dagger (i) \right)^K | - \rangle \\
| K : i,j  \rangle &=& c_K \left(\Gamma^\dagger \left(i,j \right) \right)^K | - \rangle, \\
& \cdots &    
\end{array}
\right.
\label{eq:state}
\end{eqnarray}   
with $K \le N$ 
while $c_K$ is taken by convention equal to $(K!)^{-1/2}$.
Note that, the state introduced in Eq. (\ref{eq:pbcsstate}) corresponds to the special situation where $K=N$ 
and no pair has been been removed. From these states, we define a set of coefficients from the overlaps:
\begin{eqnarray}
\left\{
\begin{array} {ccc}
I_K &=& K! ~\langle  K | K \rangle   \\
I_K(i)&=&  K! ~\langle  K: i | K : i \rangle   \\
I_K(i,j)&=&  K! ~\langle  K: i ,j | K : i,j \rangle  \\
&\cdots&   \label{eq:over}
\end{array}
\right.
\end{eqnarray}    
Using the fact that $(b^\dagger_j)^2 =0$, due to the fermionic nature of the particles, the different operators 
verify:
\begin{eqnarray}
\left(\Gamma^\dagger \left(i_1,\cdots,i_\kappa \right) \right)^K & = &  
\left(\Gamma^\dagger \left(i_1,\cdots,i_\kappa, j \right)\right)^K \nonumber \\
&+& K x_j  \left(\Gamma^\dagger \left(i_1,\cdots,i_\kappa, j \right)\right)^{K-1}.
\end{eqnarray}
This property leads to specific relationships between the states defined above and their 
overlaps. For instance:
\begin{eqnarray}
\left\{
\begin{array} {ccc}
I_{K}&=& I_{K}(i) + K |x_i|^2 I_{K-1}(i),  \\
\\
I_{K}(i) &=& I_{K}(i,j) + K |x_j|^2 I_{K-1}(i,j).  \\
& \cdots & 
\end{array}
\right.
\label{eq:rec}
\end{eqnarray}
These recurrence relations have been  recently used to solve numerically 
VAP \cite{San08} and will be at the heart of the present work to design a new functional 
for pairing. 

\subsection{Energy as an explicit functional of  $\{ x_i \}$}

Since the PBCS state is written as a functional of the parameter set $\{ x_i \}$,
expectation values of any operators can a priori be expressed as a functional of this set. Here, an illustration is given for the occupation probabilities and correlation matrix elements.
 
Using the states defined in Eq. (\ref{eq:state}), expectation values of operators 
entering into Eq. (\ref{eq:nici0}) can be expressed as
\begin{eqnarray}
\langle  N |a^\dagger_i a_i | N \rangle &=& |x_i|^2 \langle  N-1 : i  | N-1 : i \rangle \nonumber \\
\langle  N |b^\dagger_i b_j | N \rangle &=& x_i^* x_j \langle  N-1 : i | N-1 : j \rangle. \label{eq:pb}
\end{eqnarray} 
We then deduce that both occupation numbers and correlation components can be 
expressed in terms of ratios between the different coefficients introduced in Eqs. (\ref{eq:over}):
\begin{eqnarray}
\left\{
\begin{array} {lll}
n_i &=& \displaystyle  N |x_i|^2 \frac{I_{N-1}(i)}{I_N},   \\
C_{ij} &=&  \displaystyle N x^*_i x_j \frac{I_{N-1}(i,j)}{I_N} ~~{\rm for} ~~ (i \neq j),  \\
\end{array}
\right. 
\label{eq:nicipbcs}
\end{eqnarray}
while for $i = j$, $C_{ii} = \displaystyle n_i(1-n_i)$.
Overlaps entering in $n_i$ and $C_{ij}$ can be directly expressed as a functional of
$\{x_i\}$. Indeed, a direct development of $(\Gamma^\dagger)^K$ in (\ref{eq:pbcsstate})
gives:
\begin{eqnarray}
|  N \rangle &=& c_K 
  \sum_{(i_1,\cdots,i_N)}^{\ne} x_{i_1} \cdots x_{i_N} b_{i_1}^\dagger \cdots b_{i_N}^\dagger | - \rangle,\nonumber \\
&=& K! c_K  \sum_{i_1 <\cdots < i_K \le \Omega}^{\ne} x_{i_1} \cdots x_{i_N} b_{i_1}^\dagger \cdots b_{i_N}^\dagger | - \rangle,\nonumber
\end{eqnarray} 
where $\sum_{(i_1,\cdots,i_N)}^{\ne}$ is used to insist on the fact that the summation 
is made only for indices different from each others.
From this expression, it is straightforward to see that
\begin{eqnarray}
I_K &=&   \sum^{\neq}_{(i_1, \cdots ,i_{K})} |x_{i_1}|^2 \cdots |x_{i_{K}}|^2. \label{eq:elempol}
\end{eqnarray}
In a similar way, the following expressions can be deduced:
\begin{eqnarray}
\left\{
\begin{array}{ccc}
I_K(i) &=& \sum^{\neq}_{(i_1, \cdots ,i_{K}) \neq i} |x_{i_1}|^2 \cdots |x_{i_{K}}|^2  \\
\\
I_K(i,j) &=& \sum^{\neq}_{(i_1, \cdots ,i_{K}) \neq (i,j)} |x_{i_1}|^2 \cdots |x_{i_{K}}|^2\nonumber \\
\nonumber \\
& \cdots & \nonumber
\label{eq:ikijdef}
\end{array}
\right. 
\end{eqnarray} 
Note that, these expressions also suggest additional recurrence relation between the overlap:
\begin{eqnarray}
\left\{
\begin{array}{ccc}
I_K &=& \sum_{i}  |x_{i}|^2 I_K(i)  \\
\\
I_K(i) &=& \sum_{j \neq i}  |x_{j}|^2 I_K(i,j)   \\
& \cdots & 
\label{eq:rec2}
\end{array}
\right. 
\end{eqnarray} 
For completeness, additional properties are given in appendix \ref{sec:appendPBCS}.
Reporting above expressions into (\ref{eq:nicipbcs}), both $n_i$ and $C_{ij}$, and consequently the energy, take the form 
of an explicit functional of $\{ x_i \}$. This functional turns out to be too complex for a direct practical use unless one 
can take advantage of the different recurrence relation to estimate the desired quantities \cite{San08}.
   
\subsection{Energy as an implicit functional of $\{ n_i \}$}

The possibility to write the energy as a functional of natural orbitals and occupation probabilities is far from 
being trivial. Strictly speaking, Gilbert theorem \cite{Gil75} holds for systems bound by 
an external potential. It could however be extended to self-bound systems with the
introduction of Legendre multiplier technique \cite{Lie83}. In practice, 
such a technique is useful when the energy can first be written as a functional 
of the single-particle energies through some preliminary approximations 
(see for instance \cite{Pap07,Ber08}). In general, the existence of occupation number 
functional as well as its form is not straightforward. Here, we give a proof of principle
that the energy estimated with a PBCS trial wave can indeed be written as such 
a functional. {Since all quantities can be written as a functional of the $\{x_i\}$, it is sufficient to prove that
these parameters can in turn be put as a function of the $\{n_i\}$ set.}

Starting from the expression of $n_i$ and taking advantage of (\ref{eq:rec2}), we first 
obtain: 
\begin{eqnarray}
n_i  &=& N \sum_{j\ne i} |x_i|^2 |x_j|^2 \frac{I_{N-2}(i,j)}{I_{N}}.
\end{eqnarray}
Then, using the following recurrence relations 
\begin{eqnarray}
\left\{
\begin{array}{ccc}
I_{N-1}(i)&=& I_{N-1}(i,j) + (N-1) |x_j|^2 I_{N-2}(i,j) \\
\\
I_{N-1}(j)&=& I_{N-1}(i,j) + (N-1) |x_i|^2 I_{N-2}(i,j),
\end{array}
\right. \nonumber
\end{eqnarray}
which are valid for any $i \neq j$, we see that: 
\begin{eqnarray}
\left\{
\begin{array}{ccc}
I_{N-1}(i,j)  &=& \displaystyle \frac{|x_j|^2I_{N-1}(j) - |x_i|^2 I_{N-1}(i)}{|x_j|^2 -|x_i|^2}  \\
\\
I_{N-2}(i,j)  &=& \displaystyle \frac{1}{N-1}\frac{ I_{N-1}(i) - I_{N-1}(j)}{|x_j|^2 -|x_i|^2}, 
\end{array}
\right.
\label{eq:sumsub}
\end{eqnarray} 
from which we deduce 
\begin{eqnarray}
n_i (N-1)  &=&  \sum_{j\ne i}  |x_i|^2 |x_j|^2 \frac{ |x_j|^2  n_i -  |x_i|^2  n_j }{|x_j|^2 -|x_i|^2}.
\end{eqnarray}
Eventually, it can be transformed as:
\begin{eqnarray}
N (1-n_i) &=& \sum_{j\ne i} (n_j - n_i) \frac{ |x_j|^2} { |x_j|^2 - |x_i|^2}. \label{eq:dftxi}
\end{eqnarray}
This expression holds for any single-particle state $i$. This set of coupled equations between occupation 
numbers and $\{x_i\}$ is of particular interest for the present discussion. Indeed, given a set of occupation numbers $n_i$, one 
could a priori deduce the values of the $x_i$ through 
these secular equations. 
This shows that these parameters are implicit functional of the occupation probabilities (see also discussion 
in section \ref{sec:application}).

\subsection{Energy as an explicit functional of $\{ n_i \}$}

In this section, we discuss the main objective of the present work, i.e. to 
provide an explicit functional of the occupation probabilities. The strategy that is followed here 
is to use the BCS case as a guidance (see Appendix \ref{sec:bcs}). In that case, there is a direct and simple
relation between $|x_i|^2$ and $n_i$ already given in Eq. (\ref{eq:nixibcs}). Let us first see how this relation can be 
generalized in the PBCS case. 

Using the first equation of (\ref{eq:rec}) for $K=N$ and reporting in the denominator appearing in $n_i$, leads to 
\begin{eqnarray}
n_i &=& \frac{|x_i|^2}{|x_i|^2 +  \alpha_N (i)}, \label{eq:nialpha}
\end{eqnarray} 
where we have introduced the notation $\alpha_N (i) \equiv I_N(i)/(N I_{N-1}(i))$. This expression can easily 
be inverted and compared to (\ref{eq:nixibcs}). In the PBCS case, we have:
\begin{eqnarray}
|x_i|^2 &=& \left(\frac{n_i}{ 1 -n_i} \right)  \alpha_N(i). \label{eq:xi}
\end{eqnarray}
Therefore, we see that the BCS limit is recovered if $\alpha_N(i) = 1$ and that all the physics beyond the ordinary BCS or 
HFB theories is contained in its deviation from one. This could also be seen by expressing the correlation in terms of $n_i$ 
and $\alpha_N(i)$. Reporting Eq. (\ref{eq:sumsub}) into (\ref{eq:nicipbcs}), leads to
\begin{eqnarray}
C_{ij} &=&  \left\{ \begin{array}{lll}
\displaystyle n_i(1-n_i) ~~{\rm for} ~~ (i = j), \\
\\
\displaystyle x^*_i x_j \frac{n_j - n_i}{|x_j|^2- |x_i|^2} ~~{\rm for} ~~ (i \neq j)
\end{array} \right.
.
\label{eq:cijtemp}
\end{eqnarray}
Taking advantage of (\ref{eq:xi}) and using the short-hand notation $\alpha_i \equiv \alpha_N(i)$, finally gives (for $i\neq j$)
\begin{eqnarray}
C_{ij}  &=& \sqrt{n_i (1-n_i) n_j (1-n_j) \alpha_i \alpha_j}  \nonumber \\
&& \times \frac{n_i -n_j}{ n_i (1-n_j) \alpha_i - n_j (1-n_i) \alpha_j }. \label{eq:cijpbcs_alpha}
\end{eqnarray}
In the limit $\alpha_i =1$, the BCS functional $C_{ij}  = \sqrt{n_i (1-n_i) n_j (1-n_j)}$ is recovered.
More generally, it is shown that any of the following quantities, defined through:
\begin{eqnarray}
\alpha_K (i_1, \cdots ,i_\kappa) &=& \frac{1}{K}\frac{I_{K}(i_1, \cdots ,i_{\kappa})} {I_{K-1}(i_1, \cdots ,i_{\kappa})}, \nonumber\\
\end{eqnarray}
identify with $1$ in the BCS limit (see appendix \ref{sec:bcs}). 

\subsubsection{$1/N$ expansion beyond the BCS theory}

Since the BCS theory identifies to PBCS in the large $N$ limit, it is reasonable to seek for a correction 
to $\alpha_K (i_1, \cdots ,i_\kappa) = 1$ written as a $1/N$ expansion. Such an expansion 
can be obtained thanks to the relation: 
\begin{eqnarray}
\alpha_{K}(i_1, \cdots, i_K ) =& &\nonumber \\  \frac{1}{K} \sum_{j\ne(i_1, \cdots, i_K )  }^{\neq}   |x_j|^2 \hspace{-0.25cm} && \hspace{-0.25cm} \frac{\alpha_{K-1} (i_1, \cdots, i_K ,j) }{|x_j|^2 + \alpha_{K-1} (i_1, \cdots, i_K,j)},  \label{eq:hierar}
\end{eqnarray}
connecting $\alpha_{K}$  and $\alpha_{K-1}$ terms. This expression can be derived using (\ref{eq:rec}) and (\ref{eq:rec2}).
Due to the presence of a $1/K$ prefactor in this relation, any correction of order $1/(K-1)$ in $\alpha_{K-1}$ will appear as 
as an order $1/K(K-1)$ in $\alpha_{K}$. As an illustration, assuming that $\alpha_{N-1}(i,j) \simeq 1$ as in BCS, leads to:
\begin{eqnarray}
\alpha_{N}(i) &\simeq&  \frac{1}{N} \sum_j   \frac{ |x_j|^2  }{|x_j|^2 + 1 } \simeq \frac{1}{N} \sum_j   n_j \nonumber \\
&=& \frac{1}{N} (N-n_i) = 1 - \frac{1}{N} n_i,
\end{eqnarray}
that appears as the first order correction in $(1/N)$ to the BCS case. Similarly, we can obtain:
\begin{eqnarray}
\alpha_{N-1}(i,j) &\simeq&  \frac{1}{N-1} (N - n_i - n_j) \nonumber \\
\alpha_{N-2}(i,j) &\simeq&  \frac{1}{N-2} (N - n_i - n_j - n_k) \nonumber \\
&\cdots&  \nonumber
\end{eqnarray}
\begin{widetext}
Higher order corrections in $\alpha_N(i)$ can be obtained by including more and more terms 
in the expansion of all $\alpha_K$ (with $K < N$). This technique has been used in \cite{Lac10} 
to get the expansion:
\begin{eqnarray}
\alpha_N(i) &=& 1 - \frac{1}{N} n_i 
+ \frac{1}{N(N-1)}  \sum_{j \neq i} n_j^2 [ 1  -  (n_i + n_j)  ]
+ \frac{1}{N(N-1)(N-2)}  \sum_{(k,j) \neq i}^{\neq} n_j^2 n^2_k  \left[2 -  (n_i  + n_j +n_k) \right] +\cdots \nonumber\\
 \label{eq:explicit1}
\end{eqnarray}  
\end{widetext}
which corresponds to $\alpha_N(i)$
written as an explicit functional of the occupation numbers.
Note that, additional terms tested numerically as negligible and appearing at the second (or higher) 
order approximation are omitted here. This functional 
can then be injected into (\ref{eq:cijpbcs_alpha}) leading to an explicit functional of $x_i$ in 
terms of the $n_i$. Accordingly, expectation values of any operators becomes also a functional 
of the projected state occupation numbers.   

This approximation has been tested numerically in ref.  \cite{Lac10} 
and has shown a rapid convergence in the strong coupling limit. However, for small coupling (HF limit), a slow convergence 
was found. 
Indeed, assuming that $n_i \rightarrow 1$ for the $N$ pairs, we deduce, for one of the occupied state:
\begin{eqnarray}
\sum_{j \neq i} n_j^2 [ 1  -  (n_i + n_j)  ] & \rightarrow & -(N-1) \nonumber \\
\sum_{(k,j) \neq i}^{\neq} n_j^2 n^2_k  \left[2 -  (n_i  + n_j +n_k) \right] & \rightarrow & -(N-1)(N-2) \nonumber \\
&\cdots& \nonumber
\end{eqnarray}
Therefore, in this limit, all contributions to any order will participate to the same extend and  sum-up
to give $\alpha_N(i) = (1-n_i)$ leading finally to a correlation given by:
\begin{eqnarray}
C_{ij}  \rightarrow \sqrt{n_i n_j}. 
\end{eqnarray}
This form, which has been proposed using a completely different strategy in electronic system  \cite{Csa00}, will never be properly 
described by the BCS functional. From the discussion above, difficulties in the application of the present functional might also 
be anticipated. Indeed, since all terms in the expansion should be kept, the functional becomes rather complicated and its application might become rapidly intractable.
   
\subsubsection{Re-summation of the $1/N$ expansion and simplified functional}

The price to pay to correctly describes the weak coupling limit is to keep all orders in the expansion 
presented above. This basically shows that the $1/N$ expansion approach starting from the BCS 
approximation is not appropriate in that case. To overcome this difficulty a simplified functional 
can be found using the following approximation in Eq. (\ref{eq:explicit1}), 
\begin{eqnarray}
&&\frac{1}{N(N-1)}  \sum_{j \neq i}  \rightarrow   
\frac{1}{N^2}  \sum_{j} ,  \nonumber \\
&&\frac{1}{N(N-1)(N-2)}  \sum_{(k,j) \neq i}^{\neq}   \rightarrow \frac{1}{N^3}  \sum_{j k}  \cdots \nonumber
\end{eqnarray} 
while keeping all terms in this expansion. This approximation leads to a simple linear dependence 
of the $\alpha_i$ coefficient with respect to the occupation numbers $n_i$:
\begin{eqnarray}
\alpha_i = a_0 - a_1 n_i , \label{eq:linear}
\end{eqnarray}
where $a_0$ and $a_1$ are given by the expressions:
\begin{eqnarray}
a_1 &=&    \frac{1}{N} \left( 1+  s_2 + s_2^2  + 
\cdots + s_2^{N-1}   \right) \nonumber \\
&=&   \frac{1}{N}  \frac{1 - s_2^N}{1 - s_2} 
\label{eq:a1}
\end{eqnarray}  
and 
\begin{eqnarray}
a_0 &=&    1 + \frac{(s_2-s_3)}{N}  
 \left( 1+ 2  s_2 + \cdots +(N-1)   s_2^{N-2}   \right)  \nonumber \\
      & = & 1  + (s_2-s_3)  \frac{\partial a_1}{\partial s_2} , \label{eq:a0}
\end{eqnarray}  
and where the moments $s_p = \displaystyle \frac{1}{N} \sum_i (n_i)^p$ have been used. Reporting expression (\ref{eq:linear}) in correlation matrix elements Eq. (\ref{eq:cijpbcs_alpha}) gives the simple form (for $i\neq j$)
\begin{eqnarray}
C_{ij}  &=& \sqrt{n_i (1-n_i) n_j (1-n_j) }  \nonumber \\
&& \times \frac{\sqrt{ \left( a_0 - a_1 n_i \right) \left(  a_0 - a_1 n_j\right)}}{ a_0 - a_1 \left(n_i + n_j - n_i n_j \right) }.
\label{eq:funcfinal} \\
&=& {\cal C}(n_i,n_j) \nonumber
\end{eqnarray}
The functional (\ref{eq:funcfinal}) together with (\ref{eq:a1}-\ref{eq:a0}) represent the main result of this article. 
We can already anticipate some advantages of this functional (i) In the Hartree-Fock limit $s_p = 1$ for all $p>1$. Accordingly, 
$a_0 = a_1 = 1$ and we recover the HF functional quoted above, i.e. $C_{ij} = \sqrt{n_i n_j}$. (ii) The BCS limit is also easily 
identified in (\ref{eq:funcfinal}) by taking the limit $a_0 = 1$ and $a_1 = 0$. 
The net result of our approach is that the energy introduced in Eq. (\ref{eq:functotC}) that was 
originally written as a functional of the density and correlations in the projected state 
becomes now a functional of the one-body density matrix components only. In practice, such a functional 
approach should be solved by minimizing (\ref{eq:dmft}) where the energy now reads:
\begin{eqnarray}
{\cal E}_{\rm SR} [ \{\varphi_i\} , \{n_i\}]
& = & \sum_{i} t_{i} \, n_i
      + \frac{1}{2} \sum_{ij} \bar{v}^{\rho\rho}_{iijj} \,
        n_{i} n_{j} 
      \nonumber \\
&   & 
      + \frac{1}{4} \sum_{i\ne j} \bar{v}^{C}_{i\bar i j \bar j} \,
       {\cal C}(n_i,n_j)  , \nonumber \\
 &  & 
      + \frac{1}{4} \sum_{i} \bar{v}^{C}_{i\bar i i \bar i} \,
       n_i (1-n_i), \nonumber    
\end{eqnarray}   
First applications of this functional can be found in ref. \cite{Lac10} illustrating the predicting power of the functional for energies and occupations probabilities. In numerical implementation, sequential quadratic programming leads to very good convergence
at any coupling and/or large particle number.   
Below, the new functional is further illustrated and benchmarked.   
 

\section{Application}
\label{sec:application}

We consider here a system of $A$ particles interacting through the pairing 
Hamiltonian of the form \cite{Ric66a,Ric66,Ric65} 
\begin{eqnarray}
H = \sum_{i>0} \varepsilon_i (a^\dagger_i a_i + a^\dagger_{\bar i} a_{\bar i}) - \frac{g}{2} \sum_{i,j }  a^\dagger_i 
a^\dagger_{\bar i} a_{\bar j} a_{j},  \label{eq:hamilt}
\end{eqnarray} 
where $\bar i$ denotes the time-reversed state of $i$, both associated with single-particle energy $\varepsilon_i$. 
The total single-particle Hilbert space size is assumed to be $\Omega = 2A$. This Hamiltonian 
can be solved exactly numerically by making use of the so-called Richardson equations. 
First test of the functional have been made with this model Hamiltonian for an even particle number 
$A=2N$ where $N$ denotes the number of pairs and for equidistant single-particle levels, the so-called "picked fence"
Hamiltonian. Here, we will further illustrate some of the aspects of the new functional in that case, and extend the application
to even systems ($A=2N+1$) and/or non-equidistant levels.

\subsection{Illustration in the picked fence Hamiltonian}

Here, we first consider the special case of equidistant single-particle levels with a level spacing denoted by $\Delta \varepsilon$.
First, we remind that the strategy to design a functional going beyond the BCS one has been made in three steps:
(i) The parameters $\{ x_i \}$ have been first shown to be implicit functional of the $\{ n_i \}$ through 
the existence of a set of secular equation (Eq. (\ref{eq:dftxi})), (ii) Starting from the BCS prescription, 
systematic $1/N$ corrections have been proposed to get a new functional (Eq. (\ref{eq:explicit1})) (iii) Summing all orders, a 
simplified functional is then introduced (Eq. (\ref{eq:linear})). 
The step (ii) has been shown to be inadequate \cite{Lac10}, especially in the weak coupling limit. Before discussing (iii), 
the existence of secular equations as well as the uniqueness of the relation between $\{ x_i \}$ and  $\{ n_i \}$ sets of 
variational parameters is analyzed.
 
\subsubsection{Existence and uniqueness of a functional of $n_i$ }

\begin{figure}[htbq]
\includegraphics[width = 7.cm]{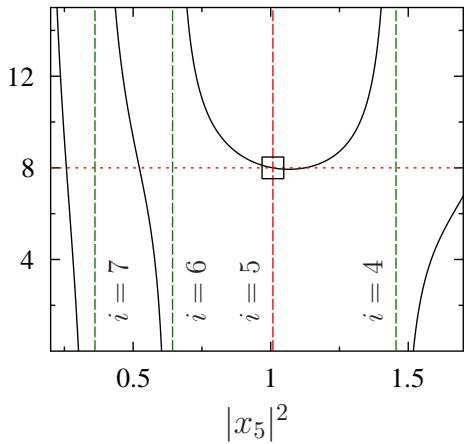}
\caption{(Color online) \label{fig1:pbcs} Graphical illustration of Eq. (\ref{eq:dftxi}) for the pairing Hamiltonian with 16 particles ($N=8$) and $g/\Delta \varepsilon = 0.22$. The solid curve corresponds to the right hand side of 
Eq. (\ref{eq:dftxi}) divided by $(1-n_i)$ as a function of $|x_5|^2$ keeping other $x_i$ fixed as well as the $n_i$. Vertical dashed lines correspond to the $|x_i|^2$ value at the minimum. The horizontal dotted line corresponds 
to $N=8$.  Equation (\ref{eq:dftxi}) is fulfilled when the dotted line crosses the solid line, which is indeed the case 
for the value of $x_5$ minimizing the functional. An open square has been added to underline the physical solution.}
\end{figure}

Eq. (\ref{eq:dftxi}) is proving the existence of a functional of the occupation number, at least an implicit one. 
A graphical illustration of Eq. (\ref{eq:dftxi}) at the PBCS energy minimum is given in 
figure \ref{fig1:pbcs}. 
The recurrence method of ref. \cite{San08} has been used to obtained the PBCS solution in this figure. 
The solid curve corresponds to the right hand side of Eq. (\ref{eq:dftxi}) divided by $(1-n_i)$ as a function of $|x_5|^2$ keeping other $x_i$ fixed as well as the $n_i$. Horizontal lines correspond to the $|x_i|^2$ value at the minimum. The horizontal dotted line corresponds to $N=8$.  Equation (\ref{eq:dftxi}) is fulfilled when the dotted line crosses the solid line, which is indeed the case for the value of $x_5$ minimizing the functional. Calculations are done for $16$ particles and a pairing constant $g/\Delta \varepsilon = 0.22$. An open square has been added to underline the physical solution. The main interest of Eq. (\ref{eq:dftxi}) is to prove that the PBCS energy can be indeed put, at least implicitly, as a functional of the occupation numbers.

This figure also illustrates that the solid line and the horizontal dotted lines cross each other several times and one may worry about the uniqueness relationship between the $\{ x_i \}$ and the $\{ n_i \}$. It should however be kept in mind that the crossing 
highlighted by the open square is the only point where Eq. (\ref{eq:dftxi}) is fulfilled {\it together with } the secular equations 
for other $|x_i|^2$. 
Indeed, starting from the recurrence relation and the expression of the occupation numbers (\ref{eq:nicipbcs}) gives
\begin{eqnarray}
n_j &=& \frac{N}{I_N} |x_j|^2 \left( I_{N-1}(i,j) + |x_i|^2 I_{N-2}(i,j) \right). \nonumber
\end{eqnarray}
Considering two states $i$ and $j$, we then have
\begin{eqnarray}
n_j - n_i &=& \frac{N}{I_N} \left( |x_j|^2 -|x_i|^2 \right) I_{N-1}(i,j). 
\end{eqnarray}
Since $I_{N-1}(i,j)/I_N > 0$, if $n_j > n_i$ then $|x_j|^2  > |x_i|^2$. This proves for instance that only 
crossing points in between $|x_4|^2$ and $|x_6|^2$ might fulfill the secular equation. A careful look at 
figure \ref{fig1:pbcs}) shows however two crossing points in this region. Let us assume that two 
solutions $|x_5|^2 = a$ and $|x_5|^2 = b$ might exist and fulfill the secular equation while keeping all other $x_i$
fixed. Using above equation, it is possible to prove that necessarily $a=b$ which finally proves that only one 
of the crossing is physical.



\subsubsection{Application of the new functional for equidistant level spacing}

\begin{figure}[htbq]
\begin{center}
\includegraphics[width = 8.5cm]{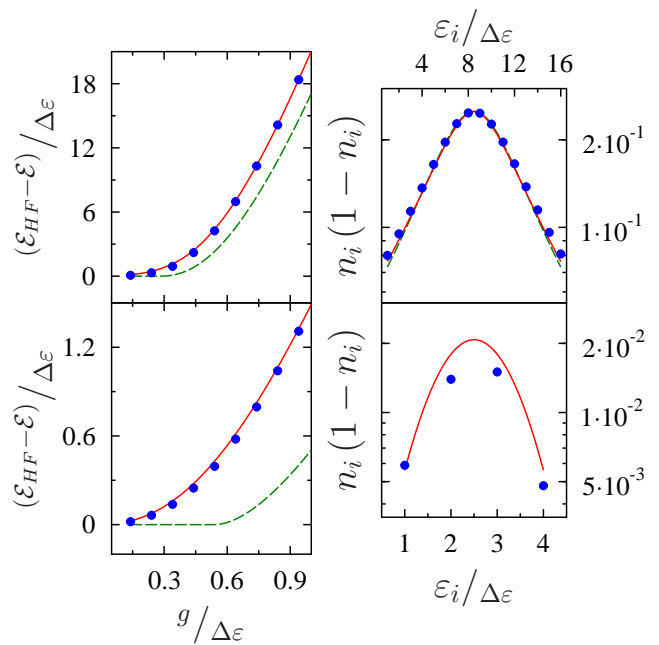} 
\end{center}
\caption{ \label{fig:evenpicked} (Color online) Evolution of the 
condensation energy for the exact (red solid line), BCS (green dash line) and new functional (blue filled circles)
obtained for the picked fence pairing Hamiltonian as a function of the coupling strength for 16 (top) and 8 (bottom) particles. 
In the right, occupation numbers of the different theories are plotted for $g/ \Delta \varepsilon=0.82$ (top) and $g/ \Delta \varepsilon=0.22$ (bottom).}
\end{figure}

We first consider the case of even systems with doubly degenerated equidistant levels. In the following, 
the condensation energy, denoted by ${\cal E}_{\rm Cond}$, defined as 
\begin{eqnarray}
{\cal E}_{\rm Cond} = {\cal E}_{HF}  - {\cal E},
\label{eq:econd}
\end{eqnarray}
where ${\cal E}_{HF} = 2\sum_{i>0} \varepsilon_i - g N$ is the Hartree-Fock (HF) energy while  ${\cal E}$ denotes the energy 
of the considered theory.  ${\cal E}_{\rm Cond}$ quantifies the predicting power of different 
approximations. An illustration of the evolution of this quantity as a function of the coupling strength
has already been given in the introduction, Fig.  (\ref{fig1:comp}). In figure \ref{fig:evenpicked},
we see that the proposed functional is almost on top of the exact result (and the exact VAP calculation). A slight difference 
is observed in the intermediate coupling regime. Similarly, occupation numbers perfectly match the 
exact ones in the strong coupling regime and slightly differ from them below the BCS 
threshold. In this regime, while BCS identifies with HF, here, occupation probabilities different 
from 1 and 0 are obtained as soon as the interaction is switched on.

\subsubsection{Critical discussion of the linear approximation (Eq. (\ref{eq:linear}))}

\begin{figure}[htbq]
\includegraphics[width = 8.5cm]{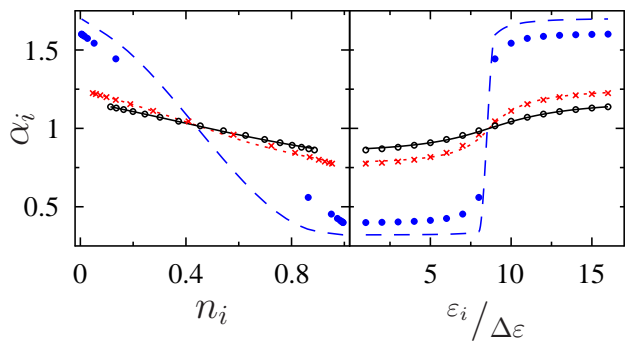}
\caption{ \label{fig:alphas} (Color online) Evolution of the coefficients $\alpha_i$ as a function of $n_i$ (left)
or $\varepsilon_i$ (right)  at the minimum of energy.
The different curves correspond to the PBCS result for $g/\Delta \varepsilon=0.32$ (dashed line), $0.64$ 
(dotted line) and $0.96$ (solid line). The corresponding results obtained with the linear approximation (Eq. 
(\ref{eq:linear})) are displayed by  filled circles, crosses and  open circles respectively. }
\end{figure}

Figure \ref{fig:alphas} shows the accuracy of the present approximation in the model case of a constant two-body interaction $g$. In this figure, the approximate $\alpha_i$ for different coupling strength $g/\Delta \varepsilon=0.32$ (filled circles), $0.64$ (crosses) and $0.96$ (open circles) are compared to the exact ones, (respectively dashed, dotted and solid lines) as a function of either the orbital probabilities (left) or single-particle energies (right) at the minimum of energy. The dependency of $\alpha_i$ obtained in the PBCS case for small coupling also shows that a simple linear approximation cannot fully grasp the physics of weak coupling. Following the same strategy as above, quadratic or cubic corrections might eventually be obtained. However, this will add complexity to the functional while the energy is already rather well reproduced. 

\subsubsection{Systematic analysis of occupation numbers}

In figure \ref{fig:evenpicked}, illustrations of occupation numbers 
obtain in different theories are shown for specific couplings. In a DMFT
framework, not only the energy should match the exact energy at the minimum but also 
the deduced one-body density matrix and a fortiori occupation numbers should also 
be identical to the exact one. To systematically compare the gain in predicting 
single-particle occupation numbers in the new functional, we have plotted the one-body entropy
\begin{eqnarray}
{\cal S} [n_i]=- \sum_i [ n_i \log(n_i) + 
(1-n_i) \log(1 - n_i) ]
\label{eq:entropy}
\end{eqnarray}
\begin{figure}[htbq]
\begin{center}
\includegraphics[width = 7.5cm]{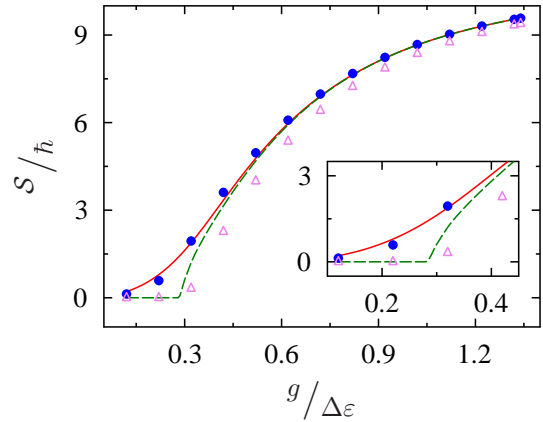}
\end{center}
\caption{ \label{fig:entropy} (Color online) Evolution of the one-body entropy for different theories as 
a function of the coupling strength for $16$ particles. The BCS (green dashed line), PAV (open violet triangles) and PBCS-functional (filled blue circles) ansatz are compared to the exact result. The inset magnifies the low $g/\Delta \varepsilon$ vicinity.}
\end{figure}in figure \ref{fig:entropy}. While PAV and BCS are unable to reproduce the exact result, especially below or in the vicinity of the threshold, the 
new functional is in close agreement with it.

\subsubsection{Simplified functional for the strong coupling regime}

One important issue from the practical point of view is the possibility to further simplify in some regime.
In particular, in the strong coupling regime, we have seen that $1/N$ perturbation starting from BCS 
rapidly converge to the exact solution. Truncation at second order of eq. (\ref{eq:explicit1}) already 
gives a very good result \cite{Lac10}. It is therefore legitimate to question whether a simpler form for 
(\ref{eq:funcfinal}) can be  found in this regime. Close to BCS, we expect $a_0 \rightarrow 1$ and 
$a_1 \rightarrow 0$ which plaid in favor of an expansion in orders of  $(a_1/a_0)$. For instance, 

\begin{eqnarray}
C_{ij} &=& \sqrt{n_i n_j(1-n_i)(1-n_j)} \nonumber\\ 
&\times& \left(1-\frac{a_1}{2a_0}   \left( n_i (1 -n_j) +n_j(1-n_i)
\right) + O \left(\frac{a_1}{a_0} \right) \right). \nonumber
\end{eqnarray}

\begin{figure}[htbq]
\begin{center}
\includegraphics[width = 7.5cm]{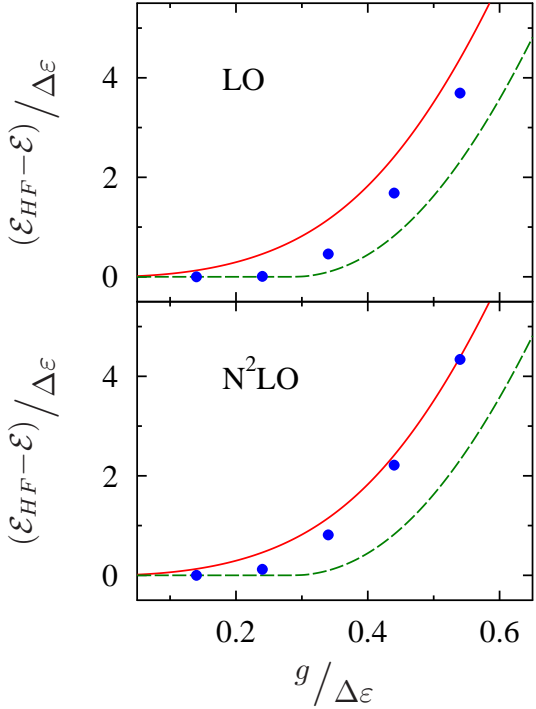} 
\end{center}
\caption{ \label{fig:funcexpand} (Color online) Comparison between leading order (top) and next to next to leading order (bottom) of the new functional (blue filled circles) for 16 particles. Evolution of the condensation energy for the exact (red solid line) and BCS (green dash line) are shown as references. 
}
\end{figure}

In Figure \ref{fig:funcexpand}, leading order (LO) [top] and next to next to leading order (N$^2$LO) [bottom] are compared as a function of the pairing strength for a typical number of particle $A=16$. 
from 0 and 1 to the exact distribution. 
It can be inferred that the full functional solution can only be recovered at low 
coupling strength when all terms of the expansion are taken into account which agrees with the previous discussion leading to resummation.

\subsubsection{Application to odd systems}

\begin{figure}[htbq]
\begin{center}
\includegraphics[width = 8.5cm]{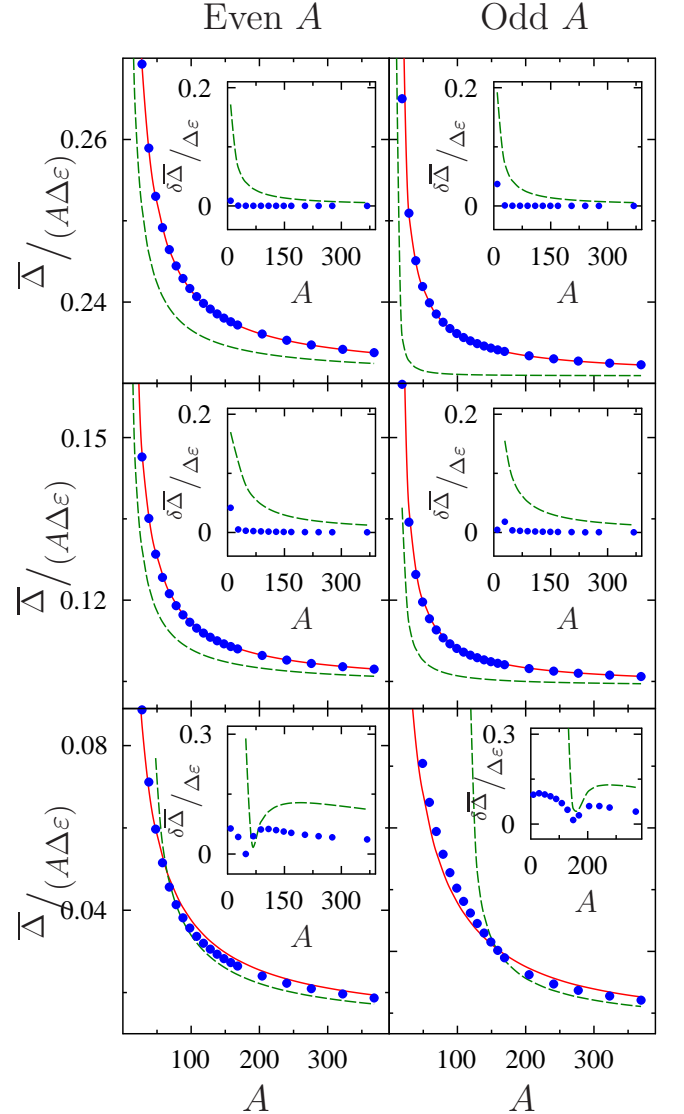}
\end{center}
\caption{ \label{fig:gap2} (Color online) Evolution of $\bar \Delta /A $ as a function of particle number $A$ for even (left)
and odd (right) systems. From top to bottom, the three different coupling constants  
$g/ \Delta \varepsilon=0.66$, $0.44$ and $0.224$ are shown. In each case, the
BCS (green dashed line) and PBCS (blue filled circles) functional theories are compared to the exact calculation (red solid line). Note that $g/ \Delta \varepsilon=0.224$ is below the BCS threshold for some values $A$ which leads to an equivalent threshold in the quantity $\bar \Delta /A $. 
In the insets, the standard deviations to the exact calculation renormalized to $1$ are compared for the different functionals.}
\end{figure}

Similarly to the BCS framework, the energy of systems with an odd 
number of particles can be obtained by using blocking techniques.
In the PBCS case, this is equivalent to consider a modified trial wave function given by
\begin{eqnarray}
| 2N + 1  \rangle &\propto&   (a^\dagger_\alpha + a^\dagger_{\bar \alpha}) 
\left(\Gamma^\dagger (\alpha) \right)^N | - \rangle \label{eq:state2}
\end{eqnarray}   
which do preserve the time-reversal symmetry of the solution. 
Here, $\{\alpha,\bar \alpha\}$ correspond to the blocked pair, and identify with the last occupied levels 
in the Hartree-Fock limit. The particle number conservation implies that occupation of the blocked states 
are kept fixed and equal to
$n_b = n_{\bar b} = 0.5$, which is nothing but the filling approximation for doubly degenerated states.
As an illustration of the odd-even effect, we define the average gap $\bar \Delta$ through the relation:
\begin{eqnarray}
\bar \Delta &=& \frac{{\cal E}_{\rm C}}{ \sum_{i\ne b} \sqrt{n_i (1-n_i)}}
\end{eqnarray}    
This quantity identifies up to a factor $1/g$ with the standard gap in the BCS limit. In figure \ref{fig:gap2}, 
the evolution of $\bar \Delta / A$  as a function of particle number $A$ is presented for different
values of the coupling strength in the exact (solid line), BCS (dashed line) and new functional (filled circles) cases. On the left, odd particle numbers are shown as compared to even ones (right), so to distinguish odd-even effect.
This figure shows that the new functional predicts well $\bar \Delta / A$ for both even and odd number of particles. Deviations at low coupling strength of the PBCS from the exact case stem from
the small discrepancies in the occupation numbers 
between those obtained in the functional formulation and the exact ones. 
The insets of Fig. \ref{fig:gap2} show the 
standard deviations from the exact calculation normalized to unity for the different 
functionals. It is worth mentioning that the same accuracy is observed for both even and 
odd systems in the case of the PBCS functional, this is in contrast with the BCS calculations. 
In the following discussion, the effect of particle number is further investigated.

\subsubsection{Accuracy of the functional with respect of particle number}

\begin{figure}[htbq]
\begin{center}
\includegraphics[width = 7.cm]{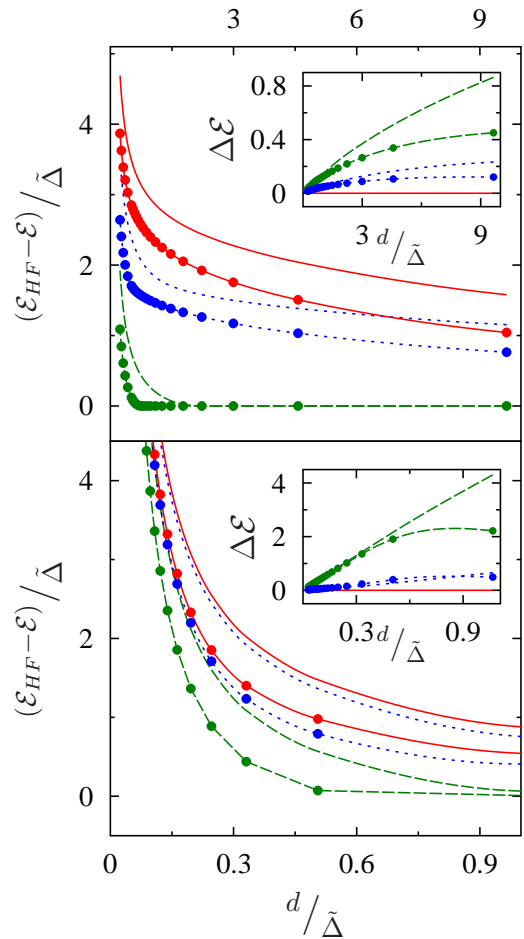}
\end{center}
\caption{ \label{fig7:en} (Color online) Condensation energy predicted by the BCS functional (green dash line) and the PBCS functional (blue filled circle) compared to the exact solution (red solid line) for varying  
$d/\widetilde{\Delta}$ and for $g/\Delta \varepsilon=0.224$ (top) and $g/\Delta \varepsilon=0.44$ (bottom). 
In each case, curves corresponding to even and odd particle number 
are shown, the latter being displayed with additional filled circles.
In the insets, relative error (in percent) on the total energy with respect to the exact solution 
made in the BCS and PBCS functionals is shown.}
\end{figure}

It is known from \cite{Bra98} that the PBCS state exhibits slight deviations from the exact solution for 
medium number of particles. Since our approach is based on a PBCS trial state, we do expect 
a similar behavior. 
To systematically address the 
quality  of the PBCS functional with respect to both the number of nucleons and the coupling strength, 
the condensation energy for odd and even systems are displayed in figure \ref{fig7:en} as a function of \cite{Bra98} $d/\widetilde{\Delta}$ 
where
\begin{eqnarray}
d/\widetilde{\Delta} \equiv \frac{2}{A} \sinh \left( 1/g \right) \nonumber
\end{eqnarray} 
for $g/\Delta \varepsilon=0.224$ (top) and $g/\Delta \varepsilon=0.44$ (bottom). In this figure, 
particle number ranging from $A=8$ (large $d$) to $A=360$ (small $d$) have been used. 
This figure illustrates the improvement of the new functional compared to BCS. It also clearly shows, that 
some deviations from the exact results persist in the new functional. It should however be kept in mind 
that the observed deviations correspond to less than $1 \%$ of errors in the total energy. 
This is illustrated in the insets of figure \ref{fig7:en}, where 
the relative error defined through
\begin{eqnarray}
\Delta {\cal E} &=& 100\frac{{\cal E} - {\cal E}_{\rm exact}}{{\cal E}_{\rm exact}}, \nonumber
\end{eqnarray}
where ${\cal E}_{\rm exact}$ is the exact energy, is displayed as a function of $d/\widetilde{\Delta}$.
As expected, error tends to zero in all cases as $A$ increases ($d \rightarrow 0$). 
For intermediate to high coupling (Fig. \ref{fig7:en}, bottom), a good agreement between the PBCS based 
functional and the exact solution is obtained, while at lower coupling strength some deviations appear. 
This results both from the approximation scheme used to design the functional 
(linear approximation for the $\alpha_i$, see \cite{Lac10}) and from the accuracy of PBCS theory itself as an approximation of the exact trial wave-function. It should indeed 
be kept in mind that the present functional is entirely based on the PBCS theory which already deviates 
from the exact solution (see for instance \cite{Fer03}). As a consequence, it could only lead to results 
which are at most equivalent to the PBCS approximation. 
From the comparison between Fig. \ref{fig7:en} (top) and ref. \cite{Sie00}, it can be inferred that deviations at low $g$ stem from (i) The deviation of PBCS result from the exact solution as $A$ increases (ii)
The additional approximations made to obtain the functional that lead to an increase of the deviation compared to PBCS as $A \rightarrow 0$. 
Nevertheless, we see from this comparison that the PBCS based functional is much more competitive than the BCS theory and 
is expected to be much easier to implement than PBCS itself.


\subsection{Application to randomly spaced levels}

As a final illustration of the functional theory application, we consider here a set of randomly spaced levels. 
Following ref. \cite{Smi96,Sie00}, an ensemble of random spectrum is generated by the central eigenvalues of a $2A \times 2A$ random matrix. 
Thus, the set of energy levels belongs to the Gaussian Orthogonal Ensemble, see ref. \cite{Meh04}. 
The renormalization proposed in ref. \cite{Smi96} is performed where
\begin{eqnarray}
\varepsilon \rightarrow 1/2\pi \left[ 4A \sin^{-1}\left( \varepsilon /\sqrt{4A}\right) - \varepsilon\sqrt{4A - \varepsilon^2}\right]
\end{eqnarray} 
so that the average level energy spacing is of the order of unity. 
\begin{figure}[htbq]
\begin{center}
\includegraphics[width = 6.cm]{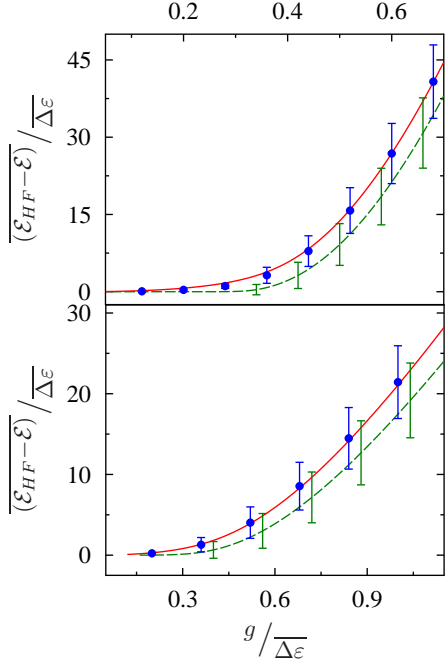}
\end{center}
\caption{ \label{fig5:erdm} (Color online) Evolution of the average condensation energy and its statistical fluctuation 
(displayed by errorbars) as a function of $g/\overline{\Delta \varepsilon}$ for the 
the PBCS based functional (blue filled circles), the BCS functional (green dot line) for $A=41$ (top) $A=16$ (bottom). 
The exact solution (red solid line) for an equidistant level spacing $\Delta \varepsilon_i$ of unit is shown as reference.}
\end{figure}
As an illustration, evolution of average condensation energy and its statistical fluctuation as a function of 
$g/\overline{\Delta \varepsilon}$ are shown 
in figure \ref{fig5:erdm}  obtained with the PBCS functional (filled circle) and the BCS functional (dot line).
Again and as expected, the new functional matches the reference result of the exact solution for an equidistant 
level spacing $\Delta \varepsilon = 1$. This last application illustrates that the method can be applied 
to systems with various level-densities.

\section{Conclusion}

In this work, quasi-particle states projected onto good particle number are used as a starting point 
to propose new functionals dedicated to pairing correlations. The properties of projected states are first 
reviewed. These properties are then used to get a functional of occupation numbers and natural 
orbitals of the trial wave-function. The new functional is benchmarked with the pairing 
Hamiltonian either with equidistant or with randomly distributed  single-particle energies for even and odd 
systems. In all cases, a very good agreement with the exact result is obtained showing great improvement 
compared to the BCS theory. Origins of the remaining deviations are discussed. 

The possibility to use a new functional accounting for particle number conservation opens new perspectives for 
the study of mesoscopic systems where pairing plays an important role. One may for instance anticipate new application 
for thermodynamics or dynamics where direct projection are too complex to provide a practical tool. In addition, this might also 
be a tool of choice avoiding recent difficulties encountered in nuclear structure studies (see for instance \cite{Lac09a}). 

\appendix

\section{Further properties of $I_K$, $I_K(i)$, ...}
\label{sec:appendPBCS}
In this appendix, properties of the overlaps defined in Eqs. (\ref{eq:over}) are further developed. The
discussion below is especially useful to make connection with recent works and between 
PBCS and BCS states. 
Using expression (\ref{eq:ikijdef}) for $I_K$ and taking advantage of  the recurrence relation (\ref{eq:rec}) gives 
\begin{eqnarray}
I_K &=&  \sum_{i} |x_i|^2 I_{K-1} - \left( N-1 \right) \sum_{i} |x_i|^4 I_{K-2} (i). \nonumber
\end{eqnarray}
In a similar way, $I_{K-2} (i)$ can be expressed in terms of $I_{K-2}$ through
\begin{eqnarray}
 I_{K-2}(i)  &=&I_{K-2} - (K-2) |x_i|^2 I_{K-3}(i), \nonumber
\end{eqnarray}
leading to 
\begin{eqnarray}
I_K &=&  \sum_{i} |x_i|^2 I_{K-1} - \left( K-1 \right) \sum_{i} |x_i|^4 I_{K-2} +  \cdots \nonumber
\end{eqnarray}
Iterating this procedure $K$ times leads to 
\begin{eqnarray}
I_{K}  &=& \sum_{n=1}^K (-1)^{n+1} \frac{\left(K-1\right)!}{\left(K-n\right)!} I_{K-n} X_n, \label{eq:newtgirar}
\end{eqnarray}
where only overlaps $I_L$ (with $L < K$ ) appear in the right hand side and where the coefficients
\begin{eqnarray}
X_n \equiv \sum_j |x_j|^{2n}, \nonumber
\end{eqnarray}
are introduced. 
According to the above expression (\ref{eq:newtgirar}), any $I_{K}$ can be written in a determinant form as
\begin{eqnarray}
I_K &=& 
\begin{array} {|cccccc|}
X_1 & 1 &0 & 0 & \cdots & 0 \\
X_2 & X_1 & 2 & 0 & \cdots & 0 \\ 
X_3 & X_2 & X_1 & 3 & \cdots & 0 \\
&& \cdots &  && \\
X_{K-1} & X_{K-2} & X_{K-3} & X_{K-4} & \cdots & (K-1) \\
X_{K} & X_{K-1} & X_{K-2} & X_{K-3} & \cdots & X_1 \\
\end{array}.\nonumber\\ \label{eq:det}
\end{eqnarray} 
The same expression has been obtained by Rowe \cite{Row91,Row01} using a completely different starting point making connection 
with elementary symmetric Schur polynomial. 
Besides this expression, similar to transformation 
between elementary symmetric polynomials and power sums $X_K$, it is worth to mention 
that other relations linking other bases of the symmetric polynomial algebra exist\cite{Mac99}. 

The same procedure can also be followed for the different quantities $I_K(i)$, $I_K(i,j)$... leading 
to a form similar  to (\ref{eq:det}) where the $X_n$ have been respectively replaced by $X_n(i)$, 
$X_n(i,j)$, ... with:
\begin{eqnarray}
X_n(i) &\equiv& \sum_{j\neq i} |x_j|^{2n} \nonumber \\
X_n(i,j) &\equiv& \sum_{k\neq (i,j)} |x_j|^{2n} \nonumber  \\
&\cdots&  \nonumber
\end{eqnarray}


\section{Guidance from the BCS theory}
\label{sec:bcs}

The BCS or HFB framework has played an important role in developing  
the new functional proposed in this work. We give here, highlights
of some aspects discussed in the text. Let us start with a state given 
by Eq.  (\ref{eq:bcsstate}). To connect with the PBCS notation, we write 
\begin{eqnarray}
| N  \rangle & \equiv & \prod_{k} \left(1 + x_k  b^\dagger_k \right) | - \rangle, \label{eq:newqp} \\
\end{eqnarray}
keeping in mind that, in the quasi-particle many-body case, the particle number $N$ is
only conserved in average and has only a meaning in the thermodynamics limit. In analogy with the PBCS case, we introduce 
the set of states $|  N-1 : i \rangle$ such that
\begin{eqnarray}
\langle  N | N \rangle &=& \langle N: i | N : i  \rangle  + |x_i|^2 
\langle  N-1 : i |  N-1 : i \rangle \nonumber \\
\langle N |a^\dagger_i a_i | N \rangle &=& |x_i|^2 \langle  N-1 : i |  N-1 : i \rangle \nonumber \\
\langle  N |b^\dagger_i b_j |  N \rangle &=& x_i^* x_j \langle  N-1 : i  | N-1 : j \rangle. \nonumber
\end{eqnarray}
Starting from (\ref{eq:newqp}), we directly see that states verifying above relations also verify:
\begin{eqnarray}
 | N : i  \rangle  &=&  | N-1 : i  \rangle = \cdots \nonumber \\
 &=&  \prod_{k \neq i} \left(1 + x_k  b^\dagger_k \right) | - \rangle. \nonumber
\end{eqnarray}
Using similar analogies between relations that hold in both PBCS and BCS case, 
we can also deduce:
\begin{eqnarray}
 | N : i,j  \rangle  &=&  | N-1 : i,j  \rangle = \cdots \nonumber \\
 &=&  \prod_{k \neq (i,j)} \left(1 + x_k  b^\dagger_k \right) | - \rangle \nonumber \\
 &\cdots&
\end{eqnarray}
Noting that the coefficient $\alpha_K$ introduced in the text also verify:
\begin{eqnarray}
\alpha_K(i)  & = & \frac{\langle  K : i |  K : i \rangle}{\langle  K-1 : i |  K-1 : i \rangle}   \nonumber \\
\alpha_K (i,j)  & = & \frac{\langle  K : i,j |  K : i,j \rangle}{\langle  K-1 : i,j |  K-1 : i,j \rangle}   \nonumber \\
&\cdots&
\end{eqnarray}
We directly see that any of these coefficients identifies to $1$ in the BCS case. 
With this in mind, let us now give some intuition on  how the BCS relation (\ref{eq:nixibcs})
can eventually be seen as a special limit of the PBCS case. Using different recurrence 
relations, it can be shown that
\begin{eqnarray}
n_{i} &=& N |x_i|^2 \frac{I_{N-1}}{I_{N}} - N(N-1) |x_i|^4 \frac{I_{N-2}}{I_{N}}  \nonumber \\
&&+ \cdots + (-1)^{N-1} N! |x_i|^{2 N } \frac{I_0}{I_{N}}  \nonumber
\end{eqnarray}
Assuming that all $\alpha_K$ are equal to $1$, gives 
\begin{eqnarray}
n_{i} &=& |x_i|^2 \left\{ 1 - |x_i|^2 + \cdots 
+ |x_i|^{2 (N-1)} \right\}, \nonumber 
\end{eqnarray}
which identifies with the BCS case, i.e. $n_i = |x_i|^2 /(1+ |x_i|^2)$ as $N \rightarrow \infty$.

\begin{acknowledgments}
We are particularly grateful to N. Sandulescu for providing us the exact Richardson and PBCS codes. We also thank Th. Duguet for helpful discussions.
\end{acknowledgments}

\bibliographystyle{apsrev4-1.bst}
\bibliography{PBCS-func}

\begin{thebibliography}{39}%
\makeatletter
\providecommand \@ifxundefined [1]{%
 \@ifx{#1\undefined}
}%
\providecommand \@ifnum [1]{%
 \ifnum #1\expandafter \@firstoftwo
 \else \expandafter \@secondoftwo
 \fi
}%
\providecommand \@ifx [1]{%
 \ifx #1\expandafter \@firstoftwo
 \else \expandafter \@secondoftwo
 \fi
}%
\providecommand \natexlab [1]{#1}%
\providecommand \enquote  [1]{``#1''}%
\providecommand \bibnamefont  [1]{#1}%
\providecommand \bibfnamefont [1]{#1}%
\providecommand \citenamefont [1]{#1}%
\providecommand \href@noop [0]{\@secondoftwo}%
\providecommand \href [0]{\begingroup \@sanitize@url \@href}%
\providecommand \@href[1]{\@@startlink{#1}\@@href}%
\providecommand \@@href[1]{\endgroup#1\@@endlink}%
\providecommand \@sanitize@url [0]{\catcode `\\12\catcode `\$12\catcode
  `\&12\catcode `\#12\catcode `\^12\catcode `\_12\catcode `\%12\relax}%
\providecommand \@@startlink[1]{}%
\providecommand \@@endlink[0]{}%
\providecommand \url  [0]{\begingroup\@sanitize@url \@url }%
\providecommand \@url [1]{\endgroup\@href {#1}{\urlprefix }}%
\providecommand \urlprefix  [0]{URL }%
\providecommand \Eprint [0]{\href }%
\providecommand \doibase [0]{http://dx.doi.org/}%
\providecommand \selectlanguage [0]{\@gobble}%
\providecommand \bibinfo  [0]{\@secondoftwo}%
\providecommand \bibfield  [0]{\@secondoftwo}%
\providecommand \translation [1]{[#1]}%
\providecommand \BibitemOpen [0]{}%
\providecommand \bibitemStop [0]{}%
\providecommand \bibitemNoStop [0]{.\EOS\space}%
\providecommand \EOS [0]{\spacefactor3000\relax}%
\providecommand \BibitemShut  [1]{\csname bibitem#1\endcsname}%
\let\auto@bib@innerbib\@empty
\bibitem [{\citenamefont {Ring}\ and\ \citenamefont {Schuck}(1980)}]{Rin80}%
  \BibitemOpen
  \bibfield  {author} {\bibinfo {author} {\bibfnamefont {P.}~\bibnamefont
  {Ring}}\ and\ \bibinfo {author} {\bibfnamefont {P.}~\bibnamefont {Schuck}},\
  }\href@noop {} {\emph {\bibinfo {title} {{The Nuclear Many-Body Problem}}}}\
  (\bibinfo  {publisher} {Springer-Verlag},\ \bibinfo {year}
  {1980})\BibitemShut {NoStop}%
\bibitem [{\citenamefont {Brink}\ and\ \citenamefont {Broglia}(2005)}]{Bri05}%
  \BibitemOpen
  \bibfield  {author} {\bibinfo {author} {\bibfnamefont {D.}~\bibnamefont
  {Brink}}\ and\ \bibinfo {author} {\bibfnamefont {R.}~\bibnamefont
  {Broglia}},\ }\href@noop {} {\emph {\bibinfo {title} {{Nuclear Superfluidity:
  pairing in finite systems}}}}\ (\bibinfo  {publisher} {Cambridge Univ.
  Press},\ \bibinfo {year} {2005})\BibitemShut {NoStop}%
\bibitem [{\citenamefont {von Delft}(2001)}]{Von01}%
  \BibitemOpen
  \bibfield  {author} {\bibinfo {author} {\bibfnamefont {J.}~\bibnamefont {von
  Delft}},\ }\href {\doibase 10.1016/S0370-1573(00)00099-5} {\bibfield
  {journal} {\bibinfo  {journal} {Phys. Rep.}\ }\textbf {\bibinfo {volume}
  {345}},\ \bibinfo {pages} {61} (\bibinfo {year} {2001})}\BibitemShut
  {NoStop}%
\bibitem [{\citenamefont {Bardeen}\ \emph {et~al.}(1957)\citenamefont
  {Bardeen}, \citenamefont {Cooper},\ and\ \citenamefont {Schrieffer}}]{Bar57}%
  \BibitemOpen
  \bibfield  {author} {\bibinfo {author} {\bibfnamefont {J.}~\bibnamefont
  {Bardeen}}, \bibinfo {author} {\bibfnamefont {L.}~\bibnamefont {Cooper}}, \
  and\ \bibinfo {author} {\bibfnamefont {J.}~\bibnamefont {Schrieffer}},\
  }\href {\doibase 10.1103/PhysRev.108.1175} {\bibfield  {journal} {\bibinfo
  {journal} {Phys. Rev.}\ }\textbf {\bibinfo {volume} {108}},\ \bibinfo {pages}
  {1175} (\bibinfo {year} {1957})}\BibitemShut {NoStop}%
\bibitem [{\citenamefont {Richardson}(1965)}]{Ric65}%
  \BibitemOpen
  \bibfield  {author} {\bibinfo {author} {\bibfnamefont {R.~W.}\ \bibnamefont
  {Richardson}},\ }\href {\doibase 10.1063/1.1704367} {\bibfield  {journal}
  {\bibinfo  {journal} {J. Math. Phys.}\ }\textbf {\bibinfo {volume} {6}},\
  \bibinfo {pages} {1034} (\bibinfo {year} {1965})}\BibitemShut {NoStop}%
\bibitem [{\citenamefont {Richardson}(1966{\natexlab{a}})}]{Ric66}%
  \BibitemOpen
  \bibfield  {author} {\bibinfo {author} {\bibfnamefont {R.}~\bibnamefont
  {Richardson}},\ }\href {\doibase 10.1103/PhysRev.144.874} {\bibfield
  {journal} {\bibinfo  {journal} {Phys. Rev.}\ }\textbf {\bibinfo {volume}
  {144}},\ \bibinfo {pages} {874} (\bibinfo {year}
  {1966}{\natexlab{a}})}\BibitemShut {NoStop}%
\bibitem [{\citenamefont {Richardson}(1966{\natexlab{b}})}]{Ric66a}%
  \BibitemOpen
  \bibfield  {author} {\bibinfo {author} {\bibfnamefont {R.}~\bibnamefont
  {Richardson}},\ }\href {\doibase 10.1103/PhysRev.141.949} {\bibfield
  {journal} {\bibinfo  {journal} {Phys. Rev.}\ }\textbf {\bibinfo {volume}
  {141}},\ \bibinfo {pages} {949} (\bibinfo {year}
  {1966}{\natexlab{b}})}\BibitemShut {NoStop}%
\bibitem [{\citenamefont {Bender}\ and\ \citenamefont {Heenen}(2003)}]{Ben03}%
  \BibitemOpen
  \bibfield  {author} {\bibinfo {author} {\bibfnamefont {M.}~\bibnamefont
  {Bender}}\ and\ \bibinfo {author} {\bibfnamefont {P.-H.}\ \bibnamefont
  {Heenen}},\ }\href {\doibase 10.1103/RevModPhys.75.121} {\bibfield  {journal}
  {\bibinfo  {journal} {Rev. Mod. Phys.}\ }\textbf {\bibinfo {volume} {75}},\
  \bibinfo {pages} {121} (\bibinfo {year} {2003})}\BibitemShut {NoStop}%
\bibitem [{\citenamefont {Stone}\ and\ \citenamefont {Reinhard}(2007)}]{Sto07}%
  \BibitemOpen
  \bibfield  {author} {\bibinfo {author} {\bibfnamefont {J.}~\bibnamefont
  {Stone}}\ and\ \bibinfo {author} {\bibfnamefont {P.}~\bibnamefont
  {Reinhard}},\ }\href {\doibase 10.1016/j.ppnp.2006.07.001} {\bibfield
  {journal} {\bibinfo  {journal} {Prog. Part. Nucl. Phys.}\ }\textbf {\bibinfo
  {volume} {58}},\ \bibinfo {pages} {587} (\bibinfo {year} {2007})}\BibitemShut
  {NoStop}%
\bibitem [{\citenamefont {BAYMAN}(1960)}]{Bay60}%
  \BibitemOpen
  \bibfield  {author} {\bibinfo {author} {\bibfnamefont {B.}~\bibnamefont
  {BAYMAN}},\ }\href {\doibase 10.1016/0029-5582(60)90279-0} {\bibfield
  {journal} {\bibinfo  {journal} {Nucl. Phys.}\ }\textbf {\bibinfo {volume}
  {15}},\ \bibinfo {pages} {33} (\bibinfo {year} {1960})}\BibitemShut {NoStop}%
\bibitem [{\citenamefont {Rodr\'{\i}guez}\ \emph {et~al.}(2005)\citenamefont
  {Rodr\'{\i}guez}, \citenamefont {Egido},\ and\ \citenamefont
  {Robledo}}]{Rod05}%
  \BibitemOpen
  \bibfield  {author} {\bibinfo {author} {\bibfnamefont {T.}~\bibnamefont
  {Rodr\'{\i}guez}}, \bibinfo {author} {\bibfnamefont {J.}~\bibnamefont
  {Egido}}, \ and\ \bibinfo {author} {\bibfnamefont {L.}~\bibnamefont
  {Robledo}},\ }\href {\doibase 10.1103/PhysRevC.72.064303} {\bibfield
  {journal} {\bibinfo  {journal} {Phys. Rev. C}\ }\textbf {\bibinfo {volume}
  {72}},\ \bibinfo {pages} {1} (\bibinfo {year} {2005})}\BibitemShut {NoStop}%
\bibitem [{\citenamefont {Anguiano}(2001)}]{Ang01}%
  \BibitemOpen
  \bibfield  {author} {\bibinfo {author} {\bibfnamefont {M.}~\bibnamefont
  {Anguiano}},\ }\href {\doibase 10.1016/S0375-9474(01)01219-2} {\bibfield
  {journal} {\bibinfo  {journal} {Nucl. Phys. A}\ }\textbf {\bibinfo {volume}
  {696}},\ \bibinfo {pages} {467} (\bibinfo {year} {2001})}\BibitemShut
  {NoStop}%
\bibitem [{\citenamefont {Dobaczewski}\ \emph {et~al.}(2007)\citenamefont
  {Dobaczewski}, \citenamefont {Stoitsov}, \citenamefont {Nazarewicz},\ and\
  \citenamefont {Reinhard}}]{Dob07}%
  \BibitemOpen
  \bibfield  {author} {\bibinfo {author} {\bibfnamefont {J.}~\bibnamefont
  {Dobaczewski}}, \bibinfo {author} {\bibfnamefont {M.}~\bibnamefont
  {Stoitsov}}, \bibinfo {author} {\bibfnamefont {W.}~\bibnamefont
  {Nazarewicz}}, \ and\ \bibinfo {author} {\bibfnamefont {P.-G.}\ \bibnamefont
  {Reinhard}},\ }\href {\doibase 10.1103/PhysRevC.76.054315} {\bibfield
  {journal} {\bibinfo  {journal} {Phys. Rev. C}\ }\textbf {\bibinfo {volume}
  {76}},\ \bibinfo {pages} {054315} (\bibinfo {year} {2007})}\BibitemShut
  {NoStop}%
\bibitem [{\citenamefont {Lacroix}\ \emph {et~al.}(2009)\citenamefont
  {Lacroix}, \citenamefont {Duguet},\ and\ \citenamefont {Bender}}]{Lac09a}%
  \BibitemOpen
  \bibfield  {author} {\bibinfo {author} {\bibfnamefont {D.}~\bibnamefont
  {Lacroix}}, \bibinfo {author} {\bibfnamefont {T.}~\bibnamefont {Duguet}}, \
  and\ \bibinfo {author} {\bibfnamefont {M.}~\bibnamefont {Bender}},\ }\href
  {\doibase 10.1103/PhysRevC.79.044318} {\bibfield  {journal} {\bibinfo
  {journal} {Phys. Rev. C}\ }\textbf {\bibinfo {volume} {79}},\ \bibinfo
  {pages} {044318} (\bibinfo {year} {2009})}\BibitemShut {NoStop}%
\bibitem [{\citenamefont {Bender}\ \emph {et~al.}(2009)\citenamefont {Bender},
  \citenamefont {Duguet},\ and\ \citenamefont {Lacroix}}]{Ben09}%
  \BibitemOpen
  \bibfield  {author} {\bibinfo {author} {\bibfnamefont {M.}~\bibnamefont
  {Bender}}, \bibinfo {author} {\bibfnamefont {T.}~\bibnamefont {Duguet}}, \
  and\ \bibinfo {author} {\bibfnamefont {D.}~\bibnamefont {Lacroix}},\ }\href
  {\doibase 10.1103/PhysRevC.79.044319} {\bibfield  {journal} {\bibinfo
  {journal} {Phys. Rev. C}\ }\textbf {\bibinfo {volume} {79}},\ \bibinfo
  {pages} {044319} (\bibinfo {year} {2009})}\BibitemShut {NoStop}%
\bibitem [{\citenamefont {Duguet}\ \emph {et~al.}(2009)\citenamefont {Duguet},
  \citenamefont {Bender}, \citenamefont {Bennaceur}, \citenamefont {Lacroix},\
  and\ \citenamefont {Lesinski}}]{Dug09}%
  \BibitemOpen
  \bibfield  {author} {\bibinfo {author} {\bibfnamefont {T.}~\bibnamefont
  {Duguet}}, \bibinfo {author} {\bibfnamefont {M.}~\bibnamefont {Bender}},
  \bibinfo {author} {\bibfnamefont {K.}~\bibnamefont {Bennaceur}}, \bibinfo
  {author} {\bibfnamefont {D.}~\bibnamefont {Lacroix}}, \ and\ \bibinfo
  {author} {\bibfnamefont {T.}~\bibnamefont {Lesinski}},\ }\href {\doibase
  10.1103/PhysRevC.79.044320} {\bibfield  {journal} {\bibinfo  {journal} {Phys.
  Rev. C}\ }\textbf {\bibinfo {volume} {79}},\ \bibinfo {pages} {044320}
  (\bibinfo {year} {2009})}\BibitemShut {NoStop}%
\bibitem [{\citenamefont {Robledo}(2010)}]{Rob10}%
  \BibitemOpen
  \bibfield  {author} {\bibinfo {author} {\bibfnamefont {L.~M.}\ \bibnamefont
  {Robledo}},\ }\href {\doibase 10.1088/0954-3899/37/6/064020} {\bibfield
  {journal} {\bibinfo  {journal} {J. Phys. G.}\ }\textbf {\bibinfo {volume}
  {37}},\ \bibinfo {pages} {064020} (\bibinfo {year} {2010})}\BibitemShut
  {NoStop}%
\bibitem [{\citenamefont {Duguet}\ and\ \citenamefont {Sadoudi}(2010)}]{Dug10}%
  \BibitemOpen
  \bibfield  {author} {\bibinfo {author} {\bibfnamefont {T.}~\bibnamefont
  {Duguet}}\ and\ \bibinfo {author} {\bibfnamefont {J.}~\bibnamefont
  {Sadoudi}},\ }\href {\doibase 10.1088/0954-3899/37/6/064009} {\bibfield
  {journal} {\bibinfo  {journal} {J. Phys. G.}\ }\textbf {\bibinfo {volume}
  {37}},\ \bibinfo {pages} {064009} (\bibinfo {year} {2010})}\BibitemShut
  {NoStop}%
\bibitem [{\citenamefont {Lacroix}\ and\ \citenamefont {Hupin}(2010)}]{Lac10}%
  \BibitemOpen
  \bibfield  {author} {\bibinfo {author} {\bibfnamefont {D.}~\bibnamefont
  {Lacroix}}\ and\ \bibinfo {author} {\bibfnamefont {G.}~\bibnamefont
  {Hupin}},\ }\href {http://arxiv.org/abs/1003.2860} {\bibfield  {journal}
  {\bibinfo  {journal} {arXiv:1003.2860}\ } (\bibinfo {year}
  {2010})}\BibitemShut {NoStop}%
\bibitem [{\citenamefont {Gilbert}(1975)}]{Gil75}%
  \BibitemOpen
  \bibfield  {author} {\bibinfo {author} {\bibfnamefont {T.}~\bibnamefont
  {Gilbert}},\ }\href {\doibase 10.1103/PhysRevB.12.2111} {\bibfield  {journal}
  {\bibinfo  {journal} {Phys. Rev. B}\ }\textbf {\bibinfo {volume} {12}},\
  \bibinfo {pages} {2111} (\bibinfo {year} {1975})}\BibitemShut {NoStop}%
\bibitem [{\citenamefont {Hohenberg}\ and\ \citenamefont {Kohn}(1964)}]{Hoh64}%
  \BibitemOpen
  \bibfield  {author} {\bibinfo {author} {\bibfnamefont {P.}~\bibnamefont
  {Hohenberg}}\ and\ \bibinfo {author} {\bibfnamefont {W.}~\bibnamefont
  {Kohn}},\ }\href {\doibase 10.1103/PhysRev.136.B864} {\bibfield  {journal}
  {\bibinfo  {journal} {Phys. Rev.}\ }\textbf {\bibinfo {volume} {136}},\
  \bibinfo {pages} {B864} (\bibinfo {year} {1964})}\BibitemShut {NoStop}%
\bibitem [{\citenamefont {Lathiotakis}\ \emph {et~al.}(2009)\citenamefont
  {Lathiotakis}, \citenamefont {Sharma}, \citenamefont {Dewhurst},
  \citenamefont {Eich}, \citenamefont {Marques},\ and\ \citenamefont
  {Gross}}]{Lat09}%
  \BibitemOpen
  \bibfield  {author} {\bibinfo {author} {\bibfnamefont {N.}~\bibnamefont
  {Lathiotakis}}, \bibinfo {author} {\bibfnamefont {S.}~\bibnamefont {Sharma}},
  \bibinfo {author} {\bibfnamefont {J.}~\bibnamefont {Dewhurst}}, \bibinfo
  {author} {\bibfnamefont {F.}~\bibnamefont {Eich}}, \bibinfo {author}
  {\bibfnamefont {M.}~\bibnamefont {Marques}}, \ and\ \bibinfo {author}
  {\bibfnamefont {E.}~\bibnamefont {Gross}},\ }\href {\doibase
  10.1103/PhysRevA.79.040501} {\bibfield  {journal} {\bibinfo  {journal} {Phys.
  Rev. A}\ }\textbf {\bibinfo {volume} {79}},\ \bibinfo {pages} {16} (\bibinfo
  {year} {2009})}\BibitemShut {NoStop}%
\bibitem [{\citenamefont {Lacroix}\ \emph {et~al.}(2004)\citenamefont
  {Lacroix}, \citenamefont {Ayik},\ and\ \citenamefont {Chomaz}}]{Lac04a}%
  \BibitemOpen
  \bibfield  {author} {\bibinfo {author} {\bibfnamefont {D.}~\bibnamefont
  {Lacroix}}, \bibinfo {author} {\bibfnamefont {S.}~\bibnamefont {Ayik}}, \
  and\ \bibinfo {author} {\bibfnamefont {P.}~\bibnamefont {Chomaz}},\ }\href
  {\doibase 10.1016/j.ppnp.2004.02.002} {\bibfield  {journal} {\bibinfo
  {journal} {Prog. Part. Nucl. Phys.}\ }\textbf {\bibinfo {volume} {52}},\
  \bibinfo {pages} {497} (\bibinfo {year} {2004})}\BibitemShut {NoStop}%
\bibitem [{\citenamefont {Sheikh}\ and\ \citenamefont {Ring}(2000)}]{She00}%
  \BibitemOpen
  \bibfield  {author} {\bibinfo {author} {\bibfnamefont {J.-A.}\ \bibnamefont
  {Sheikh}}\ and\ \bibinfo {author} {\bibfnamefont {P.}~\bibnamefont {Ring}},\
  }\href {http://linkinghub.elsevier.com/retrieve/pii/S0375947499004248}
  {\bibfield  {journal} {\bibinfo  {journal} {Nucl. Phys. A}\ }\textbf
  {\bibinfo {volume} {665}},\ \bibinfo {pages} {71} (\bibinfo {year}
  {2000})}\BibitemShut {NoStop}%
\bibitem [{\citenamefont {Rowe}\ \emph {et~al.}(1991)\citenamefont {Rowe},
  \citenamefont {Song},\ and\ \citenamefont {Chen}}]{Row91}%
  \BibitemOpen
  \bibfield  {author} {\bibinfo {author} {\bibfnamefont {D.}~\bibnamefont
  {Rowe}}, \bibinfo {author} {\bibfnamefont {T.}~\bibnamefont {Song}}, \ and\
  \bibinfo {author} {\bibfnamefont {H.}~\bibnamefont {Chen}},\ }\href {\doibase
  10.1103/PhysRevC.44.R598} {\bibfield  {journal} {\bibinfo  {journal} {Phys.
  Rev. C}\ }\textbf {\bibinfo {volume} {44}},\ \bibinfo {pages} {R598}
  (\bibinfo {year} {1991})}\BibitemShut {NoStop}%
\bibitem [{\citenamefont {Rowe}(2001)}]{Row01}%
  \BibitemOpen
  \bibfield  {author} {\bibinfo {author} {\bibfnamefont {D.}~\bibnamefont
  {Rowe}},\ }\href {\doibase 10.1016/S0375-9474(01)00588-7} {\bibfield
  {journal} {\bibinfo  {journal} {Nucl. Phys. A}\ }\textbf {\bibinfo {volume}
  {691}},\ \bibinfo {pages} {691} (\bibinfo {year} {2001})}\BibitemShut
  {NoStop}%
\bibitem [{\citenamefont {Sandulescu}\ and\ \citenamefont
  {Bertsch}(2008)}]{San08}%
  \BibitemOpen
  \bibfield  {author} {\bibinfo {author} {\bibfnamefont {N.}~\bibnamefont
  {Sandulescu}}\ and\ \bibinfo {author} {\bibfnamefont {G.}~\bibnamefont
  {Bertsch}},\ }\href {\doibase 10.1103/PhysRevC.78.064318} {\bibfield
  {journal} {\bibinfo  {journal} {Phys. Rev. C}\ }\textbf {\bibinfo {volume}
  {78}},\ \bibinfo {pages} {1} (\bibinfo {year} {2008})}\BibitemShut {NoStop}%
\bibitem [{\citenamefont {Sandulescu}\ \emph {et~al.}(2009)\citenamefont
  {Sandulescu}, \citenamefont {Errea},\ and\ \citenamefont {Dukelski}}]{San09}%
  \BibitemOpen
  \bibfield  {author} {\bibinfo {author} {\bibfnamefont {N.}~\bibnamefont
  {Sandulescu}}, \bibinfo {author} {\bibfnamefont {B.}~\bibnamefont {Errea}}, \
  and\ \bibinfo {author} {\bibfnamefont {J.}~\bibnamefont {Dukelski}},\ }\href
  {\doibase 10.1103/PhysRevC.80.044335} {\bibfield  {journal} {\bibinfo
  {journal} {Phys. Rev. C}\ }\textbf {\bibinfo {volume} {80}},\ \bibinfo
  {pages} {044335} (\bibinfo {year} {2009})}\BibitemShut {NoStop}%
\bibitem [{\citenamefont {Leggett}(2006)}]{Leg06}%
  \BibitemOpen
  \bibfield  {author} {\bibinfo {author} {\bibfnamefont {A.~J.}\ \bibnamefont
  {Leggett}},\ }\href {\doibase 10.1093/acprof:oso/9780198526438.001.0001}
  {\emph {\bibinfo {title} {{Quantum Liquids}}}},\ Vol.~\bibinfo {volume} {1}\
  (\bibinfo  {publisher} {Oxford University Press},\ \bibinfo {address}
  {Oxford, England},\ \bibinfo {year} {2006})\BibitemShut {NoStop}%
\bibitem [{\citenamefont {Lieb}(1983)}]{Lie83}%
  \BibitemOpen
  \bibfield  {author} {\bibinfo {author} {\bibfnamefont {E.~H.}\ \bibnamefont
  {Lieb}},\ }\href {\doibase 10.1002/qua.560240302} {\bibfield  {journal}
  {\bibinfo  {journal} {Int. J. Quant. Chem.}\ }\textbf {\bibinfo {volume}
  {24}},\ \bibinfo {pages} {243} (\bibinfo {year} {1983})}\BibitemShut
  {NoStop}%
\bibitem [{\citenamefont {Papenbrock}\ and\ \citenamefont
  {Bhattacharyya}(2007)}]{Pap07}%
  \BibitemOpen
  \bibfield  {author} {\bibinfo {author} {\bibfnamefont {T.}~\bibnamefont
  {Papenbrock}}\ and\ \bibinfo {author} {\bibfnamefont {A.}~\bibnamefont
  {Bhattacharyya}},\ }\href {\doibase 10.1103/PhysRevC.75.014304} {\bibfield
  {journal} {\bibinfo  {journal} {Phys. Rev. C}\ }\textbf {\bibinfo {volume}
  {75}},\ \bibinfo {pages} {1} (\bibinfo {year} {2007})}\BibitemShut {NoStop}%
\bibitem [{\citenamefont {Bertolli}\ and\ \citenamefont
  {Papenbrock}(2008)}]{Ber08}%
  \BibitemOpen
  \bibfield  {author} {\bibinfo {author} {\bibfnamefont {M.}~\bibnamefont
  {Bertolli}}\ and\ \bibinfo {author} {\bibfnamefont {T.}~\bibnamefont
  {Papenbrock}},\ }\href {\doibase 10.1103/PhysRevC.78.064310} {\bibfield
  {journal} {\bibinfo  {journal} {Phys. Rev. C}\ }\textbf {\bibinfo {volume}
  {78}},\ \bibinfo {pages} {064310} (\bibinfo {year} {2008})}\BibitemShut
  {NoStop}%
\bibitem [{\citenamefont {Cs\'{a}nyi}\ and\ \citenamefont
  {Arias}(2000)}]{Csa00}%
  \BibitemOpen
  \bibfield  {author} {\bibinfo {author} {\bibfnamefont {G.}~\bibnamefont
  {Cs\'{a}nyi}}\ and\ \bibinfo {author} {\bibfnamefont {T.}~\bibnamefont
  {Arias}},\ }\href {\doibase 10.1103/PhysRevB.61.7348} {\bibfield  {journal}
  {\bibinfo  {journal} {Phys. Rev. B}\ }\textbf {\bibinfo {volume} {61}},\
  \bibinfo {pages} {7348} (\bibinfo {year} {2000})}\BibitemShut {NoStop}%
\bibitem [{\citenamefont {Braun}\ and\ \citenamefont {von
  Delft}(1998)}]{Bra98}%
  \BibitemOpen
  \bibfield  {author} {\bibinfo {author} {\bibfnamefont {F.}~\bibnamefont
  {Braun}}\ and\ \bibinfo {author} {\bibfnamefont {J.}~\bibnamefont {von
  Delft}},\ }\href {\doibase 10.1103/PhysRevLett.81.4712} {\bibfield  {journal}
  {\bibinfo  {journal} {Phys. Rev. Lett.}\ }\textbf {\bibinfo {volume} {81}},\
  \bibinfo {pages} {4712} (\bibinfo {year} {1998})}\BibitemShut {NoStop}%
\bibitem [{\citenamefont {Fern\'{a}ndez}\ and\ \citenamefont
  {Egido}(2003)}]{Fer03}%
  \BibitemOpen
  \bibfield  {author} {\bibinfo {author} {\bibfnamefont {M.}~\bibnamefont
  {Fern\'{a}ndez}}\ and\ \bibinfo {author} {\bibfnamefont {J.}~\bibnamefont
  {Egido}},\ }\href {\doibase 10.1103/PhysRevB.68.184505} {\bibfield  {journal}
  {\bibinfo  {journal} {Phys. Rev. B}\ }\textbf {\bibinfo {volume} {68}},\
  \bibinfo {pages} {184505} (\bibinfo {year} {2003})}\BibitemShut {NoStop}%
\bibitem [{\citenamefont {Sierra}\ \emph {et~al.}(2000)\citenamefont {Sierra},
  \citenamefont {Dukelsky}, \citenamefont {Dussel}, \citenamefont {von Delft},\
  and\ \citenamefont {Braun}}]{Sie00}%
  \BibitemOpen
  \bibfield  {author} {\bibinfo {author} {\bibfnamefont {G.}~\bibnamefont
  {Sierra}}, \bibinfo {author} {\bibfnamefont {J.}~\bibnamefont {Dukelsky}},
  \bibinfo {author} {\bibfnamefont {G.}~\bibnamefont {Dussel}}, \bibinfo
  {author} {\bibfnamefont {J.}~\bibnamefont {von Delft}}, \ and\ \bibinfo
  {author} {\bibfnamefont {F.}~\bibnamefont {Braun}},\ }\href {\doibase
  10.1103/PhysRevB.61.R11890} {\bibfield  {journal} {\bibinfo  {journal} {Phys.
  Rev. B}\ }\textbf {\bibinfo {volume} {61}},\ \bibinfo {pages} {R11890}
  (\bibinfo {year} {2000})}\BibitemShut {NoStop}%
\bibitem [{\citenamefont {Smith}\ and\ \citenamefont
  {Ambegaokar}(1996)}]{Smi96}%
  \BibitemOpen
  \bibfield  {author} {\bibinfo {author} {\bibfnamefont {R.}~\bibnamefont
  {Smith}}\ and\ \bibinfo {author} {\bibfnamefont {V.}~\bibnamefont
  {Ambegaokar}},\ }\href {\doibase 10.1103/PhysRevLett.77.4962} {\bibfield
  {journal} {\bibinfo  {journal} {Phys. Rev. Lett.}\ }\textbf {\bibinfo
  {volume} {77}},\ \bibinfo {pages} {4962} (\bibinfo {year}
  {1996})}\BibitemShut {NoStop}%
\bibitem [{\citenamefont {Mehta}(2004)}]{Meh04}%
  \BibitemOpen
  \bibfield  {author} {\bibinfo {author} {\bibfnamefont {M.}~\bibnamefont
  {Mehta}},\ }\href@noop {} {\emph {\bibinfo {title} {{Random matrices}}}}\
  (\bibinfo  {publisher} {Academic Press},\ \bibinfo {year} {2004})\BibitemShut
  {NoStop}%
\bibitem [{\citenamefont {Macdonald}(1999)}]{Mac99}%
  \BibitemOpen
  \bibfield  {author} {\bibinfo {author} {\bibfnamefont {I.~G.}\ \bibnamefont
  {Macdonald}},\ }\href@noop {} {\emph {\bibinfo {title} {Bulletin of the
  American Mathematical Society}}},\ Vol.~\bibinfo {volume} {4}\ (\bibinfo
  {publisher} {Oxford University Press, USA},\ \bibinfo {year}
  {1999})\BibitemShut {NoStop}%
\end{thebibliography}%

\end{document}